\documentclass[prb,aps,twocolumn]{revtex4}
\usepackage{amsmath,epsfig,color,verbatim}
\begin{document}
\title{Electronic spectrum in high-temperature cuprate superconductors} %
\author{N.M.Plakida$^{a,b}$ and V.S. Oudovenko$^{a,c}$  }
\affiliation{$^a$Joint Institute for Nuclear Research, 141980 Dubna,  Russia\\
$^b$Max-Planck-Institut f\"ur Physik komplexer Systeme, D-01187 Dresden,
Germany\\
$^c$Rutgers University, Piscataway, New Jersey 08854, USA}

\begin{abstract}
A microscopic theory for electronic  spectrum of the CuO$_2$ plane within an
effective $p$-$d$ Hubbard model  is proposed. Dyson equation for the
single-electron Green function in terms of the Hubbard operators is derived
which is solved self-consistently for the self-energy evaluated in the
noncrossing approximation. Electron scattering on spin fluctuations induced by
kinematic interaction is described by a dynamical spin susceptibility with a
continuous spectrum. Doping and temperature dependence of electron dispersions,
spectral functions, the Fermi surface and the coupling constant $\lambda$ are
studied in the hole doped case. At low doping, an arc-type Fermi surface  and a
pseudogap in the spectral function  are observed.

\end{abstract}


\maketitle

\section{Introduction}

Recent high-resolution angle-resolved photoemission spectroscopy (ARPES) studies
revealed a complicated character of electronic structure and quasiparticle (QP)
spectra in copper oxide superconductors.  In particular, a pseudogap in the
electronic spectrum and  an arc-type Fermi surface (FS) at low hole
concentrations were revealed, a substantial wave-vector and energy dependent
renormalization of the Fermi-velocity of QP (``kinks" in the dispersion) was
observed  (see, e.g., \cite{Damascelli03,Sadovskii01,Eschrig05} and references
therein). As was originally pointed out by Anderson~\cite{Anderson87}, strong
electron correlations in cuprates  play an essential role in explaining their
normal and superconducting properties.
\par
A conventional approach in describing strong electron correlations is based on
consideration of the Hubbard model~\cite{Hubbard63}. The model has  some
advantages in comparison with the $t$-$J$ model which can be derived  from the
Hubbard model in the limit of strong correlations.  Namely, the Hubbard model
allows to study a moderate correlation limit observed experimentally in curates
and more consistently takes into account  a two-subband character of electronic
structure, in particular, a weight transfer between subbands with doping.
\par
Various methods were proposed to study electronic structure within the Hubbard
model. An unbiased method  based on  numerical simulations for finite clusters
(for a review see e.g.,~\cite{Bulut02})  precludes, however, to study subtle
features of QP spectra due to poor energy and wave-vector resolutions in small
size clusters. In analytical calculations of spectra  mean-field type
approximations are often used (for a review see~\cite{Ovchinnikov04,Mancini04})
which cannot reproduce the above mentioned effects caused by the self-energy
contributions. In the  dynamical mean field theory (DMFT) (for a review
see~\cite{Georges96,Kotliar05}) the self-energy is treated in the single-site
approximation which also unable to describe wave-vector dependent phenomena. To
overcome this flaw of DMFT, various types of the dynamical cluster theory  were
developed (for a review see~\cite{Maier04,Tremblay06}). In these methods only a
restricted wave-vector and energy resolutions can be achieved, depending on the
size of the clusters, while the physical interpretation of the origin of an
anomalous electronic structure in  numerical methods is not straightforward.
\par
To elucidate the mechanism of the pseudogap formation, the charge carriers
scattering by short-ranged (static) antiferromagnetic (AF) spin fluctuations was
considered in several analytical semi-phenomenological studies (for a review
see~\cite{Sadovskii01}). More recently, by including into the DMFT scheme an
additional momentum-dependent component of the self-energy originating from
short-range AF (or charge) correlations, the spin-fluctuation scenario of the
pseudogap formation~\cite{Sadovskii05} and  the arc-type FS~\cite{Kuchinskii05}
have been supported  (for a review see~\cite{Kuchinskii06}). At the same time,
it is important to study the effects of the charge carriers scattering by the
{\it dynamical} spin-fluctuations which are believed to be responsible for the
kink phenomenon~\cite{Eschrig05}. This can be done by considering the Dyson
equation for the single-particle Green function (GF) within the Hubbard model in
the limit of strong correlations.  For instance, calculation of electronic
spectrum within the first order perturbation theory for the self-energy has
reproduced quite accurately quantum Monte Carlo results~\cite{Krivenko05}, while
application of an incremental cluster expansion for the self-energy has enabled
to observe a kink structure in the QP spectrum~\cite{Kakehashi04}.
\par
The aim of the present paper is to develop a microscopic  theory for the
electronic spectrum in strongly correlated systems, as cuprates, which
consistently takes into account effects of electron scattering by dynamical spin
fluctuations.  For this, we have considered an effective Hubbard model reduced
form the $p$-$d$ model for the CuO$_2$ plane in cuprates. By applying the
Mori-type projection technique for the thermodynamic GF~\cite{Zubarev60} in
terms of the Hubbard operators, we derived an {\it exact} Dyson equation, as was
elaborated in our previous publications~\cite{Plakida95,Plakida99, Plakida03}. A
self-consistent  solution of the Dyson equation  with the self-energy evaluated
in the noncrossing approximation (NCA) beyond  a perturbation approach was
performed.
\par
This enabled us to calculate the dispersion and spectral functions of
single-particle excitations, the FS, and the electron occupation numbers. In
particular, we studied a hole-doped case at various hole concentrations. At low
doping, the FS reveals an arc-type shape with a pseudogap in the $(\pi, 0)$
region of the Brillouin zone (BZ). A strong renormalization effects of the
dispersion close to the Fermi energy (``kinks") are observed due to electron
scattering by dynamical AF spin fluctuations induced by kinematic interaction
generic for the Hubbard operators. Electron occupation numbers show only a small
drop at the Fermi energy. For high temperature or large hole concentrations, AF
correlations become weak and a crossover to a Fermi-liquid-like behavior is
observed.
\par
In the next Section we briefly discuss the model and  derivation of the Dyson
equation, and the self-energy calculation in the NCA.  The results of numerical
solution of the self-consistent system of equations for various hole
concentrations  and discussion  are presented in Sect.~3. Conclusion is given in
Sect.~4.

\section{General formulation}

\subsection{Effective Hubbard model and Dyson equation}

Following a cell-cluster perturbation theory (e.g.,
\cite{Plakida95,Feiner96,Yushankhai97}) based on a consideration of the original
two-band $p$-$d$ model for the CuO$_2$ layer~\cite{Emery87}  we consider an
effective two-dimensional Hubbard model for holes
\begin{eqnarray}
H &= & \varepsilon_1\sum_{i,\sigma}X_{i}^{\sigma \sigma} +
\varepsilon_2\sum_{i}X_{i}^{22} +  \sum_{i\neq
j,\sigma}\bigl\{t_{ij}^{11}X_{i}^{\sigma 0}X_{j}^{0\sigma}
 \nonumber \\
& + & t_{ij}^{22}X_{i}^{2 \sigma}X_{j}^{\sigma 2} +2\sigma
t_{ij}^{12}(X_{i}^{2\bar\sigma}X_{j}^{0 \sigma} + {\rm H.c.})\bigr\},
 \label{m1}
\end{eqnarray}
where $X_{i}^{nm} = |in\rangle\langle im|$ are the Hubbard operators (HOs) for
the four states $\,n, m=|0\rangle ,\,|\sigma\rangle, |2\rangle =|\uparrow
\downarrow \rangle $, $\sigma = \pm 1/2 = (\uparrow,\downarrow)$,
$\bar\sigma=-\sigma$. Here $\varepsilon_1=\varepsilon_d-\mu$ and
$\varepsilon_2=2\varepsilon_1+  U_{eff} $ where $\mu$ is the chemical potential.
The effective  Coulomb energy in the Hubbard model~(\ref{m1}) is the
charge-transfer  energy  $ U_{eff} = \Delta = \epsilon_p-\epsilon_d$. The
superscript $2$ and $1$ refers to the two-hole $p$-$d$ singlet subband  and the
one-hole subband, respectively. According to the cell-cluster perturbation
theory, we can take similar values for the hopping parameters in (\ref{m1}): $\,
t^{22}_{ij}  = t^{11}_{ij} = t^{12}_{ij} = t_{ij}$.  The bare electron
dispersion  defined  by the hopping parameter $ t_{ij}$ we determine by the
conventional  equation
\begin{equation}
t({\bf k}) = 4 t \, \gamma({\bf k}) + 4 t' \,\gamma'({\bf k}) + 4 t''\,
\gamma''({\bf k}) ,
 \label{m1a}
\end{equation}
where $t, \, t', \, t''\,$ are the hopping parameters for  the nearest neighbor
(n.n.) $ ( \pm a_{x}, \pm a_{y})$,  the next nearest neighbor (n.n.n.) $\pm (a_x
\pm a_y)$ and $\, \pm 2 a_{x}, \pm 2 a_{y}$ sites, respectively, and
$\gamma({\bf k})= (1/2)(\cos k_x +\cos k_y), \,  \gamma'({\bf k}) = \,\cos k_x
\cos k_y\, $  and $\gamma '' ({\bf k})= (1/2)(\cos 2 k_x +\cos 2 k_y) $ (the
lattice constants $ a_{x}= a_{y}$ equal to unity).  To get a physically
reasonable value for the charge-transfer gap for the conventional value of $ t
\simeq 0.4$~eV we take $\;\Delta = U_{eff} = 8\, t \simeq 3.2$~eV. The bare
bandwidth is $W = 8 t \simeq U_{eff}$ which shows that the effective $p$-$d$
Hubbard model~(\ref{m1}) corresponds to the strong  correlation limit. In what
follows,  the energy will be measured in unit of $\,t \,$ with $\varepsilon_d =
0$ in $\varepsilon_1$. The chemical potential $\mu$ depends on the average {\it
hole} occupation number
\begin{equation}
  n =  1 + \delta
    = \langle\, \sum\sb{\sigma}  X\sb{i}\sp{\sigma \sigma} +
      2  X\sb{i}\sp{22} \rangle .
\label{m2}
\end{equation}
The HOs entering~(\ref{m1}) obey the completeness relation $\, X_{i}^{00} +
X_{i}^{\uparrow \uparrow} + X_{i}^{\downarrow \downarrow}
 + X_{i}^{22} = 1 \,$
which rigorously preserves the constraint of no double occupancy of any quantum
state $|in\rangle$ at each lattice site $i$. Due to the projected character of
the HOs, they have complicated commutation relations
$\,\left[X_{i}^{\alpha\beta}, X_{j}^{\gamma\delta}\right]_{\pm}=
\delta_{ij}\left(\delta_{\beta\gamma}X_{i}^{\alpha\delta}\pm
\delta_{\delta\alpha}X_{i}^{\gamma\beta}\right)$, which results in the so-called
{\it kinematic interaction}. The upper sign here refers for the  Fermi-like HOs
like $X_{i}^{0\sigma}$ and the lower sign is for the Bose-like ones, like the
spin or number operators.
\par
To discuss the electronic structure within the  model~(\ref{m1}), we introduce a
thermodynamic matrix Green function (GF)~\cite{Zubarev60}
\begin{eqnarray}
{\hat G}_{ij\sigma}(t-t')&=& \langle\langle \hat X_{i\sigma}(t)\! \mid \! \hat
X_{j\sigma}^{\dagger}(t')\rangle\rangle
\nonumber \\
  &=&  -i\theta(t-t')\langle \{ \hat X_{i\sigma}(t)\, , \, \hat
X_{j\sigma}^{\dagger}(t')\}\rangle ,
 \label{m4}
\end{eqnarray}
in terms of the two-component operators
$\, \hat X_{i\sigma} = \left( \begin{array}{c} X_i^{\sigma 2} \\
X_i^{0\bar\sigma} \end{array} \right)$ and $\, \hat X_{i\sigma}^{\dagger} =
(X_{i}^{2\sigma}\,\, X_{i}^{\bar\sigma 0}) $. To calculate the GF~(\ref{m4}), we
apply the Mori-type projection technique by writing equations of motion for the
Heisenberg operators in the form:
\begin{equation}
   {\hat Z}_{i \sigma} = [\hat X_{i\sigma},\, H]=
\sum_{j}\,\hat{\varepsilon}_{ij\sigma} \hat X_{j\sigma} +
 {\hat Z}_{i \sigma}^{(ir)} ,
 \label{m5}
 \end{equation}
where the {\it irreducible} $\hat Z $-operator is determined by the
orthogonality condition:
\begin{equation}
\langle\{{\hat Z}_{i \sigma}^{(ir)},\, {\hat X}_{j\sigma}^{\dagger}\}\rangle =
 \langle {\hat Z}_{i\sigma}^{(ir)}\,{\hat X}_{j\sigma}^{\dagger} +
{\hat X}_{j\sigma}^{\dagger}\, {\hat Z}_{i\sigma}^{(ir)} \rangle = 0\, .
 \label{m5a}
 \end{equation}
This defines  the frequency matrix
\begin{equation}
\hat{\varepsilon}\sb{ij} =  \langle\{[\hat X_{i\sigma},H],\hat
X_{j\sigma}^{\dagger}\}\rangle \; \hat{Q}^{-1} ,
  \label{m6}
\end{equation}
where  $ \hat{Q} =\langle\{\hat X_{i\sigma},\hat X_{i\sigma}^{\dagger}\}\rangle
=
\left(  \begin{array}{cc} Q\sb{2} & 0 \\
      0 & Q\sb{1} \end{array} \right) $. The weight factors
 $\, Q\sb{2} = \langle X\sb{i}\sp{22} + X\sb{i}\sp{\sigma\sigma}\rangle
 =  n/2 \,$ and  $\, Q\sb{1} = \langle X\sb{i}\sp{00} +
X\sb{i}\sp{\bar\sigma \bar\sigma} \rangle = 1-Q\sb{2}\, $ in a paramagnetic
state  depend only on the hole occupation number (\ref{m2}). The frequency
matrix (\ref{m6}) determines the QP spectra within the generalized mean field
approximation (MFA). The corresponding zero-order GF in MFA reads:
\begin{equation}
    {\hat G}^{\, 0}_\sigma({\bf k},\omega) =
   \Bigl(\omega \hat{\tau}_{0} -
   \hat{\varepsilon}({\bf k}) \Bigr)^{-1}\hat{Q},
  \label{m7}
\end{equation}
where $\,\hat\tau_{0}$ is the unity matrix and we introduced the frequency
matrix (\ref{m6}) in the ${\bf k}$-representation $\hat{\varepsilon}({\bf k})$.
By differentiating the many-particle GF $\,\langle\langle \hat
Z_{i\sigma}^{irr}(t)\! \mid \! \hat X_{j\sigma}^{\dagger}(t')\rangle\rangle\, $
over the second time $t'$ and applying the same projection procedure as in
(\ref{m5}) we derive the Dyson equation in the form~\cite{Plakida95}
\begin{equation}
 {\hat G}_\sigma({\bf k},\omega)^{-1}=
 {\hat G}_{\sigma}^{\, 0}({\bf k}, \omega)^{-1} -
 {\hat \Sigma}_{\sigma}({\bf k},\omega).
  \label{m8}
\end{equation}
Here the self-energy matrix $\,{\hat \Sigma}_{\sigma}({\bf k},\omega)$ is
determined by a {\it proper } part (which  have no single zero-order GF) of the
many-particle GF in the form
\begin{equation}
{\hat \Sigma}_{\sigma}({\bf k}, \omega) = {\hat{Q}}^{-1} \langle\!\langle {\hat
Z}_{\sigma}^{(ir)} \!\mid\!  {\hat Z}_{\sigma}^{(ir)\dagger}
\rangle\!\rangle^{(prop)}_{{\bf k}, \omega}\;{\hat{Q}}^{-1}.
 \label{m9}
\end{equation}
The equations~(\ref{m7}) -- (\ref{m9}) provide an exact representation for the
GF~(\ref{m4}). However, to calculate it one has to use an approximation for the
self-energy matrix~(\ref{m9}) which describes inelastic scattering of electrons
on spin and charge fluctuations.
\par
It is important to point out that  in the Hubbard model~(\ref{m1})  electron
interaction with spin- or charge fluctuations are induced by the kinematic
interaction with the coupling constants equal to the original hopping
parameters, as has been already pointed out by  Hubbard~\cite{Hubbard63}. For
instance, the equation of motion for the operator $X\sb{i}\sp{\sigma 2} $ reads
\begin{eqnarray}
  id \, X\sb{i}\sp{\sigma 2}/dt &= &[X\sb{i}\sp{\sigma 2}, H] =
   (\varepsilon_1 + \Delta) X\sb{i}\sp{\sigma2}
\nonumber \\
  &+& \sum\sb{l\ne i,\sigma '}\! \left( t\sb{il}\sp{22}
    B\sb{i\sigma\sigma '}\sp{22} X\sb{l}\sp{\sigma ' 2} -
   2 \sigma t\sp{21}\sb{il} B\sb{i\sigma\sigma '}\sp{21}
    X\sb{l}\sp{0\bar\sigma '} \right)
\nonumber \\
 &-& \sum\sb{l\ne i} X\sb{i}\sp{02} \left( t\sp{11}\sb{il}
    X\sb{l}\sp{\sigma0} + 2 \sigma t\sp{21}\sb{il}
    X\sb{l}\sp{2 \bar\sigma} \right),
\label{m10}
\end{eqnarray}
where  $B\sb{i\sigma\sigma'}\sp{\alpha\beta}$ are Bose-like operators describing
the number (charge) and spin fluctuations:
\begin{eqnarray}
  B\sb{i\sigma\sigma'}\sp{22} & = & (X\sb{i}\sp{22} +
   X\sb{i}\sp{\sigma\sigma}) \delta\sb{\sigma'\sigma} +
   X\sb{i}\sp{\sigma\bar\sigma} \delta\sb{\sigma'\bar\sigma}
\nonumber\\
 & = & ( N\sb{i}/2 + S\sb{i}\sp{z})
\delta\sb{\sigma'\sigma} +
    S\sb{i}\sp{\sigma}\delta\sb{\sigma'\bar\sigma},
  \label{m11}\\
  B\sb{i\sigma\sigma'}\sp{21} & = & ( N\sb{i}/2 +
   S\sb{i}\sp{z}) \delta\sb{\sigma'\sigma} -
   S\sb{i}\sp{\sigma} \delta\sb{\sigma'\bar\sigma}\, .
\nonumber
\end{eqnarray}
Therefore, in the Hubbard model (\ref{m1}), contrary to spin-fermion models
where electron interaction  with spin- or charge fluctuations are specified  by
fitting coupling constants~\cite{Eschrig05}, this interaction is fixed by the
hopping parameters.

\subsection{Mean-Field Approximation}
\label{MFA}

The single-particle excitations in MFA are defined by the frequency matrix
(\ref{m6}). By using  equations of motion like (\ref{m10}),  we get the
following  energy spectrum for holes in two  subbands
\begin{eqnarray}
{\varepsilon}_{1, 2} ({\bf k})& = & ({1}/{2}) [\omega_{2} ({\bf k}) + \omega_1
({\bf k})] \mp({1}/{2}) \Lambda({\bf k}),
 \nonumber\\
  \Lambda({\bf k}) &= &    \{[\omega_{2} ({\bf k})
  - \omega_1 ({\bf k})]^2 + 4 W({\bf k})^2 \}^{1/2},
\label{n1}
\end{eqnarray}
where the original excitation spectra in the Hubbard subbands and the
hybridization parameter are
\begin{eqnarray}
{\omega}_1({\bf k})& = &
 4  t\,\alpha_{1} \gamma({\bf k})
 + 4 t'\,\beta_{1}\gamma'({\bf k}) - \mu,
\nonumber \\
{\omega}_2({\bf k})& = &
 4  t\,\alpha_{2} \gamma({\bf k})
 + 4 t'\,\beta_{2} \gamma'({\bf k}) + \Delta - \mu,
\nonumber \\
 W({\bf k}) & = &  4  t\,\alpha_{12} \gamma({\bf k})
 + 4 t' \,\beta_{12} \gamma'({\bf k}).
  \label{n2}
\end{eqnarray}
where we omitted  $t''$ contribution in (\ref{m1a}) and introduced the
renormalization parameters  $\, \alpha_{1(2)}= Q_{1(2)}[ 1 +
{C_{1}}/{Q^2_{1(2)}}], \,
 \beta_{1(2)} = Q_{1(2)}[ 1 + {C_{2}}/{Q^2_{1(2)}}]\,$,
$\,  \alpha_{12}= \sqrt{Q_{1}Q_{2}}[ 1 - {C_{1}}/{Q_{1}Q_{2}}] ,\,
 \beta_{12} =  \sqrt{Q_{1}Q_{2}}[ 1 -{C_{2}}/{Q_{1}Q_{2}}]\,$. As
in the Hubbard I approximation, we neglect number  fluctuations $\langle \delta
N_i \delta N_j \rangle_{(i \neq j)}\,$  but take into account contributions from
the spin correlation functions for the n.n. and the n.n.n. sites:
\begin{equation}
C_{1} = \langle {\bf S}_i{\bf S}_{i\pm a_{x}/a_{y}} \rangle, \quad C_{2} =
\langle {\bf S}_i{\bf S}_{i\pm a_{x}\pm a_{y}} \rangle  .
 \label{n3}
\end{equation}
The renormalization of the QP spectra (\ref{n1}), (\ref{n2}) caused by strong
spin correlations in the underdoped region results in suppression of the n.n.
hopping which changes the shape of the spectra and  reduces the bandwidth. For
instance, if we consider the limiting case of the long-range AF N\'eel  state
with the n.n. correlation function $C_{1} \simeq - 1/4 $ at half-filling, $Q_{1}
= Q_{2} = 1/2$, we obtain $\alpha_{1(2)} = 0$. This  results in complete
suppression of the n.n. hopping and transformation of  the spectra (\ref{n2}) to
the n.n.n. hopping
 $ \propto t' \gamma'({\bf k})$ as was discussed in~\cite{Plakida95}.
\par
 For the diagonal components of
the zero-order  GF (\ref{m7})  we have
\begin{equation}
 G_{11(22)}^{\, 0}({\bf k},\omega)=
  \frac{Q_{1(2)}\,[1 - b({\bf k})]}{\omega - {\varepsilon}_{1(2)}({\bf k})}
 +  \frac{Q_{1(2)}\,b({\bf k})}{\omega - {\varepsilon}_{2(1)} {\bf(k)}} ,
 \label{n4}
\end{equation}
where the parameter
\begin{equation}
 b({\bf k}) = \frac{{\varepsilon}_{2} ({\bf k}) - \omega_{2}({\bf k})}
 {{\varepsilon}_{2} ({\bf k}) - {\varepsilon}_{1} ({\bf k})}=
 \frac{1}{2} -  \frac{\omega_{2}({\bf k})
- \omega_1({\bf k})}{ 2 \Lambda({\bf k})}
 \label{n5}
\end{equation}
determines the contribution due to the hybridization.

\subsection{Self-energy Corrections}

Dyson equation (\ref{m8}) for the GF is convenient to write in the form
\begin{equation}
 {\hat G}_\sigma({\bf k},\omega)= \left(\omega
\hat{\tau}_{0} - \hat {\varepsilon}({\bf k}) - \tilde{\Sigma}_{\sigma}({\bf k},
\omega)\right)^{-1} \hat {Q},
  \label{s1}
\end{equation}
where the self-energy reads
\begin{equation}
\tilde{\Sigma}_{\sigma}({\bf k}, \omega) =\langle\!\langle {\hat
Z}_{\sigma}^{(ir)} \!\mid\!  {\hat Z}_{\sigma}^{(ir)\dagger}
\rangle\!\rangle^{(prop)}_{{\bf k}, \omega}\;{\hat{Q}}^{-1} . \label{s2}
\end{equation}
To make the problem tractable, we can neglect in the self-energy matrix
(\ref{s2}) the off-diagonal components $\tilde{\Sigma}_{12,\sigma}({\bf
k},\omega)$  in comparison with the hybridization parameters $\, W({\bf k})\, $
in (\ref{n2}). This enables us to write the diagonal components of the full
GF~(\ref{s1}) in the form similar to (\ref{n4}):
\begin{eqnarray}
 {\hat G}_{11(22)}({\bf k},\omega)  =
  \frac{Q_{1(2)} \, [1 - b({\bf k})]}
 {\omega - {\varepsilon}_{1(2)}({\bf k}) -
 \tilde{\Sigma}_{11(22)}({\bf k},\omega)}
\nonumber \\
+  \frac{Q_{1(2)} \, b({\bf k})}
 {\omega - {\varepsilon}_{2(1)}({\bf k})
  - \tilde{\Sigma}_{22(11)}({\bf k}, \omega)}\, .
 \label{s3}
\end{eqnarray}
Here the hybridization parameters $b({\bf k})$ are determined  by the formula
similar to (\ref{n5}) which gives an accurate approximation for a small doping
at $n \sim 1 $.
\par
Now we calculate the self-energy~(\ref{s2})  in the non-crossing (NCA) or the
self-consistent Born approximation (SCBA) by neglecting vertex renormalization.
As follows from the equation of motion  (\ref{m10}),  the ${\hat
Z}_{\sigma}^{(ir)} $ operators determined by (\ref{m5})  are essentially a
product of a Fermi-like $ X_{j}(t)$ and Bose-like $ B_{i}(t)$ operators.  In
SCBA, the propagation of these excitations of different types in the
many-particle GF in~(\ref{s2}) are assumed to be independent of each other.
Therefore, they can be decoupled  in the time-dependent correlation functions
for lattice sites $\,(i \neq j,\, l \neq m)$ as follows
\begin{equation}
  \langle B_{i}(t) X_{j}(t) B_{l} X_{m} \rangle
\simeq \langle X_{j}(t) X_{m} \rangle \langle B_{i}(t) B_{l} \rangle.
 \label{s4}
\end{equation}
Using the spectral representation for these correlation functions, we obtain the
following formula for the diagonal self-energy components
$\,\tilde{\Sigma}_{11(22)}({\bf k},\omega)=
 {\Sigma}({\bf k},\omega)\,$ which are the same for two subbands:
\begin{eqnarray}
 {\Sigma}({\bf k},\omega)
 &= &\frac{1}{  N } \,\sum\sb{\bf q}
   \int\limits\sb{-\infty}\sp{+\infty} \!\!{\rm d}z
    K(\omega,z|{\bf q},{\bf k - q})
 \nonumber \\
 & \times  &  (- {1}/{ \pi })\,\mbox{Im}\,
   [{G}_{1}({\bf q},z)  + {G}_{2}({\bf q},z) ] ,
 \label{s8}
\end{eqnarray}
where the corresponding  subband  GFs are:
\begin{equation}
{G}_{1 (2)}({\bf q},\omega) =  \frac{1}
 {\omega - {\varepsilon}_{1 (2)} ({\bf q})-
 {\Sigma}({\bf q},\omega)} \,  .
 \label{s9}
\end{equation}
 The kernel of the integral equation (\ref{s8}) has the following form:
\begin{eqnarray}
&& K(\omega,z|{\bf q},{\bf k - q })= | t({\bf q})|\sp{2}\; \frac{1}{2\pi}
 \int\limits\sb{-\infty}\sp{+\infty}\;
   \frac{{\rm d}\Omega}{\omega - z - \Omega}
\nonumber\\
 &\times&[ \tanh ({z}/{2T}) +
 \coth ({\Omega}/{2T})] \,  \mbox{Im} \,
 \chi\sb{sc}({\bf k -q},\Omega), \quad
\label{s6}
\end{eqnarray}
where the interaction is defined by the hopping parameter $t({\bf q}) $
(\ref{m1a}). The spectral density of bosonic excitations is determined by the
dynamic susceptibility of the Bose-like operators $ B_{i}(t)$ in ~(\ref{s4}) --
the spin and number (charge) fluctuations:
\begin{eqnarray}
 \chi\sb{sc}({\bf q},\omega) =
 -  [ \langle\!\langle {\bf S\sb{q} | S\sb{-q}} \rangle\!\rangle\sb{\omega}
  + ({1}/{4}) \langle\!\langle \delta N\sb{\bf q} | \delta N\sb{-\bf q}
   \rangle\!\rangle\sb{\omega} ] \,,
\label{s7}
\end{eqnarray}
where we introduced the commutator GF for the spin ${\bf S \sb{q}} $ and the
number $\delta N_{\bf q} =  N_{\bf q} - \langle N_{\bf q} \rangle$ operators.
\par
Thus we obtain a self-consistent system of equations for the GFs ~(\ref{s9}) and
the self-energy (\ref{s8}). A similar system of equations was obtained within
the composite operator method~\cite{Krivenko05}. In comparison with the $t$-$J$
model studied by us in~\cite{Plakida99}, for the Hubbard model (\ref{m1}) we
have two contributions in the self-energy ~(\ref{s8}) determined by the two
Hubbard subbands, while in the $t$-$J$ model only one subband is considered.
However, depending on the position of the chemical potential,  a substantial
contribution to the self-energy comes only from the GF of those subband which is
close to the Fermi energy.  A contribution from the GF of another subband which
is far from the Fermi energy,  is suppressed due to a large charge-transfer
energy $ \Delta $ in the denominator of those GF. Neglecting the latter
contribution, we  obtain a self-consistent system of equations for one  GF close
to the Fermi energy and the corresponding self-energy function similar to the
$t$-$J$ model ~\cite{Plakida99}.

\section{Results and discussion}

\subsection{Self-consistent system of equations}
\label{system}

To solve the system of equations for  the self-energy (\ref{s8}) and the GFs
(\ref{s9}) we should  specify a model for the spin-charge susceptibility
(\ref{s7}). Below  we take into account  only the spin-fluctuation contribution
$\, \chi_{s}({\bf q},\omega) = - \langle \langle {\bf S}_{q}\mid {\bf S}_{-q}
\rangle \rangle _{\omega}$ for which we adopt a model suggested in numerical
studies~\cite{Jaklic95}
\begin{eqnarray}
&&     {\rm Im}\, \chi_{s}({\bf q},\omega+i0^+) =
 \chi_{s}({\bf q}) \; \chi_{s}^{''}(\omega)
\nonumber\\
 & = &  \frac {\chi_0}{1+ \xi^2 (1+ \gamma({\bf q}))} \;  \tanh
\frac{\omega}{2T} \frac{1}{1+(\omega/\omega_{s})^2}\, .
 \label{r1}
\end{eqnarray}
The ${\bf q}$-dependence in $\chi_{s}({\bf q})$ is determined by the  AF
correlation length $\xi$ which doping dependence is defined below. The static
susceptibility $\chi_0$ at the AF wave vector ${\bf Q = (\pi,\pi)}$ is fixed by
the normalization condition:
 \begin{eqnarray}
 && \langle {\bf S}_{i}^2\rangle =
 \frac{1}{N}\sum_{i} \langle {\bf S}_{i}{\bf S}_{i} \rangle
  \nonumber \\
  &=&
 \frac{1}{\pi} \, \int\limits_{-\infty}^{+\infty }
  \frac{dz}{\exp{(z/T)} - 1}  \chi_{s}^{''}(z)\;
  \frac{1}{N} \sum_{\bf q} \chi_{s}({\bf q}) ,
 \label{r2}
\end{eqnarray}
which gives the following value for  this  constant:
\begin{equation}
 \chi_{0}=
  \frac{2}{\omega_{s}}\,\langle {\bf S}_{i}^2\rangle \left \{\frac{1}{N}
   \sum_{\bf q} \frac{1}
{1+\xi^2[1+\gamma({\bf q})]} \right\}^{-1} .
 \label{r3}
\end{equation}
In (\ref{r2}) we introduced $\langle {\bf S}_{i}^2\rangle =
 3 \langle S^z_{i}\,S^z_{i}\rangle= ({3}/{4})
 \langle (1 - X_i^{00} - X_i^{22})\rangle \simeq
({3}/{4})(1- |\delta|)$ where at the hole doping $ \delta \simeq \langle
X_i^{22}\rangle$, while  at the electron doping $ \delta \simeq  - \langle
X_i^{00}\rangle $.
\par
The spin correlation functions (\ref{n3}) in the single-particle excitation
spectra (\ref{n1}) in MFA  are defined  by equations
\begin{equation}
C_1 =  \frac{1}{N} \sum_{\bf q}\, C_{\bf q}\, \gamma({\bf q}), \quad
 C_2 = \frac{1}{N} \sum_{\bf q} \, C_{\bf q}\, \gamma'({\bf q}).
 \label{r4}
\end{equation}
The static correlation function $\, C_{\bf q}\, $ can be calculated from the
same model (\ref{r1}) as follows
\begin{equation}
C_{\bf q} = \langle {\bf S}_{\bf q}{\bf S}_{-\bf q} \rangle =
 \frac{C(\xi)} {1+\xi^2[1+ \gamma({\bf q})]} \, ,
 \label{r5}
\end{equation}
where the factor $C(\xi) = \chi_{0}\,( {\omega_{s}}/{2})$.
\par
To specify the doping dependence of the AF correlation length $\xi (\delta)$ at
low temperature, we fit the correlation function $\, C_1 \,$  calculated from
(\ref{r4}) to the numerical results of an exact diagonalization  for finite
clusters ~\cite{Bonca89}. The values of the  AF correlation length, calculated
values of $C_2$ and  the correlation function $\, C(\xi) = \langle {\bf S}_{\bf
q}{\bf S}_{-\bf q} \rangle\,$ at the AF wave-vector ${\bf q = Q} = (\pi, \pi)\,$
are given in Table~\ref{Table1}.
\begin{table}
\centering \caption{Static spin correlation  functions (\ref{r4}), $C(\xi)$
(\ref{r5}) and the AF correlation length $\xi$ in (\ref{r1}) at various hole
concentrations $ n = 1 + \delta$}
\label{Table1}      
 \begin{tabular}{lrrrrrr}
\noalign {\bigskip}
 \noalign{\smallskip}\hline
\noalign{\smallskip}
 \quad $\delta =$  \quad  & 0.03  & 0.05  & 0.10 & 0.15  &  0.20 & 0.30\\
 \noalign{\smallskip} \hline\\
 \quad $C_1$    \quad &-0.36 &-0.26 &-0.21 &-0.18  &-0.14 &-0.10  \\
 \quad $C_2$    \quad & 0.27 & 0.16 & 0.11 & 0.09  & 0.06 & 0.04  \\
 \quad $C(\xi)$ \quad &22.0  & 5.91 & 3.58 & 2.67  & 1.93 & 1.40  \\
 \quad $ \xi$   \quad & 8.0  & 3.40 & 2.50 & 2.10  & 1.70 & 1.40  \\
\noalign{\smallskip}
 \hline\noalign{\smallskip}
\end{tabular}
\end{table}
\par
To perform numerical calculations, we introduce  the imaginary frequency
representation for the GF (\ref{s9}):
\begin{equation}
{G}_{1 (2)}({\bf q},i\omega_n) =  \frac{1}
 {i\omega_n - {\varepsilon}_{1 (2)} ({\bf q})-
 {\Sigma}({\bf q},i\omega_n)} \,  .
 \label{r6}
\end{equation}
where $ i\omega_{n}=i\pi T(2n+1), \; n = 0,\pm 1, \pm 2,$ ... . For the
self-energy (\ref{s8}) we obtain the following representation:
\begin{eqnarray}
 {\Sigma}({\bf k}, i\omega_{n}) &=&
 - \frac{T}{N}\sum_{\bf q}
 \sum_{m}   [{G}_{1}({\bf q}, i\omega_{m}) +
{G}_{2}({\bf q}, i\omega_{m})]
 \nonumber \\
&\times & \lambda({\bf q, k-q} \mid i\omega_{n}-i\omega_{m})\, .
 \label{r7}
\end{eqnarray}
The interaction function  is given here by the equation
\begin{eqnarray}
\lambda({\bf q, k-q} \mid i\omega_{\nu}) = -  |t({\bf q})|^{2} \,
 \chi_{s}({\bf k-q}) \;
  F_{s}(i\omega_{\nu}),
  \label{r8}
\end{eqnarray}
where  the spectral function:
\begin{equation}
F_s(\omega_\nu)=\frac{1}{\pi} \;  \int_{0}^{\infty}\frac {2x dx}{x^2 +
(\omega_\nu/\omega_s)^2} \,  \frac{1}{1+x^2} \,
 \tanh \frac {x\,\omega_{s} }{2T}.
 \label{r9}
\end{equation}
Let us compare the self-consistent system of equations for  the GF (\ref{r6})
and the self-energy (\ref{r7}) with  results of other theoretical approaches. In
our theory based on the HO technique we start from the two-subband
representation for the GF~(\ref{m4}) which rigorously takes into account strong
electron correlations determined by the Coulomb energy $U_{eff}$. This results
in the Mott gap at large $U_{eff}$ (see below) as in the DMFT. On the other
hand, the kinematic interaction, generic to HOs, induces the electron scattering
by spin (charge) dynamical fluctuations (\ref{s7}) which are responsible for the
pseudogap formation as in the two-particle self-consistent approach (TPSC)
~\cite{Vilk95,Tremblay06}  or the  model of short-range static spin (charge)
fluctuations -- the $\Sigma_{\bf k}$-model~\cite{Sadovskii01}.
\par
To prove this, let us consider  the classical limit  for the self-energy
(\ref{r7}) by taking into account only the zero Matsubara frequency
$i\omega_{\nu} =0$ in the interaction (\ref{r8}) which gives $i\omega_{m}=
i\omega_{n}$ in  (\ref{r7}). In the limit of large  AF correlation length $\xi
\gg 1 $ the static spin susceptibility $\chi_{s}({\bf q})$ in~(\ref{r1}) shows a
sharp peak close to the AF wave-vector ${\bf Q} = (\pi,\pi)$ and can be expanded
over the small wave-vector ${\bf p = q - Q}$:
\begin{equation}
 \chi_{s}({\bf  q})\simeq \, \frac {\chi_0 }
 {1 + \xi^2 \,{\bf p}^2}
  \simeq \frac {A}{\kappa^{2}+{\bf p}^2} .
 \label{ar1}
\end{equation}
where we introduced $\kappa = \xi^{-1} $ and took into account that the constant
(\ref{r3}) $\chi_0 \simeq A\,\xi^2$  with $\, A=
  ({8\pi}/{\omega_{s}}) \langle {\bf S}_{i}^2 \rangle
  [\ln(1 + 4\pi \, \xi^2)]^{-1}$ for the square lattice.
In this limit we get the following equation for the self-energy
 (\ref{r7}):
 \begin{eqnarray}
 &&{\Sigma}({\bf k}, i\omega_{n}) \simeq |g({\bf k-Q})|^{2}
 \, \frac{T}{N}\sum_{\bf p} \,\frac {1}
{{\kappa^{2}+ p^2}}
  \nonumber \\
 && \times  [{G}_{1}({\bf k - Q -p}, i\omega_{n}) +
{G}_{2}({\bf k-Q -p}, i\omega_{n})], \quad
 \label{ar7}
\end{eqnarray}
where the effective interaction
\begin{equation}
|g({\bf q})|^{2} = A\,|t({\bf  q})|^{2}\, F_s(0) .
 \label{ar8}
\end{equation}
Expanding the QP energy $\,\varepsilon_{1 (2)}({\bf k-Q -p}) \simeq
\varepsilon_{1 (2)}({\bf k-Q}) - {\bf p \cdot v}_{1 (2), \bf k-Q}\,$ we obtain
for the GFs in (\ref{ar7}) the following representation:
\begin{eqnarray}
&&{G}_{1(2)}({\bf k - Q -p}, i\omega_{n})
 \simeq \{i\omega_n - {\varepsilon}_{1 (2)} ({\bf k- Q})
\nonumber \\
&+&
 {\bf p \cdot v}_{1 (2), \bf k-Q}-
  \Sigma({\bf k -Q},i\omega_n)\}^{-1} .
  \label{ar6}
\end{eqnarray}
The system of equations for the GFs (\ref{ar6}) and the self-energy (\ref{ar7})
is similar to those one derived in the TPSC  approach~\cite{Vilk95}) and the
$\Sigma_{\bf k}$-model~\cite{Sadovskii01} apart from the interaction function
and the two-subband system of equations. In our approach the vertex (\ref{ar8})
is determined by the hopping parameter $|t({\bf k-Q})|^2 $, while in the TPSC
and the $\Sigma_{\bf k}$-model the coupling constant is induced by the Coulomb
scattering, e.g., in~\cite{Kuchinskii06} $g^2 = U^2 ( \langle n_{i\uparrow} n_{i
\downarrow}\rangle / n^2 )\langle {\bf S}_{i}^2\rangle /3$. However, the values
of these vertices are close: the averaged over the BZ value $
\langle\sqrt{|t({\bf k})|^2}\rangle_{{\bf k}} \sim 2t$ is comparable   with the
coupling constant $ g \leq 2t$ used in~\cite{Sadovskii05}. In the spin-fermion
model the self-energy is also determined by spin-fluctuations (see,
e.g.,~\cite{Eschrig05}) with the coupling constant fitted from ARPES experiments
$g \sim 0.7$~eV$\sim 2t$ of the same order. As in the TPSC theory, in the limit
$\xi \rightarrow \infty $  the AF gap $\Delta_{AF}({\bf k}) \propto |t({\bf
k-Q})|^{2} $ in the QP spectra emerges in the subband located at the Fermi
energy. This result readily follows from the self-consistent equations for the
GF (\ref{r6}) with the self-energy (\ref{ar7}) where in the right-had side GF
(\ref{ar6}) is taken at ${\bf p} =0$. Thus, in our approach the pseudogap
formation is mediated by the AF short-range order similar to TPSC theory and the
model of short-range static spin fluctuations in the generalized
DMFT~~\cite{Kuchinskii06}.
\par
In the next sections we consider the results of self-consistent calculations of
the GFs (\ref{r6}) and the self-energy (\ref{r7}) in the hole doped case for
various hole concentration $\delta = n - 1 > 0$. In Sects.~\ref{DA} -- \ref{SE}
the calculations were performed at temperature $T = 0.03 t \simeq 140$~K and $T
= 0.3 t$ for $\Delta =8 t,\, t \simeq 0.4$~eV  and $t' = - 0.3t $. Several
results are reported for $\Delta =4 t,\, t' = - 0.13t, \, t'' = 0.16t $ in
Sect.~\ref{D4}. For the spin-fluctuation energy in (\ref{r1}) we take $\omega_s
= 0.4t$. The AF correlation length $\xi(\delta)$ and the static correlation
functions $C_1, C_2$ in (\ref{n3}) are defined in Table~\ref{Table1}.

\subsection{Dispersion and spectral functions}
\label{DA}

In ARPES measurements  and QMC simulations the spectrum of single-electron
excitations is determined by the spectral function $\, {A}_{(el)}({\bf k},
\omega)
 = {A}_{(h)}({\bf k}, -\omega) \,$. The spectral function for holes
can be written as follow:
\begin{eqnarray}
 &&{A}_{(h)}({\bf k}, \omega)= - \frac{1}{\pi}\,{\rm Im}\,
\langle\langle a_{{\bf k}\sigma}\, |
 \, a_{{\bf k}\sigma}^{\dag} \rangle\rangle_{\omega + i0^+}
\nonumber \\
& = &  [Q_1 +  P({\bf k})]{A}_{1}({\bf k}, \omega)
 +[Q_2 -  P({\bf k})]{A}_{2}({\bf k}, \omega).
 \label{r10}
\end{eqnarray}
Here we introduced  for the  hole annihilation $ a_{{\bf k}\sigma}$ and creation
$ a_{{\bf k}\sigma}^{\dag}$ operators the definition in terms of the Hubbard
operators $\,  a_{{\bf k}\sigma} = X\sb{i}\sp{0\sigma} +
  2\sigma X\sb{i}\sp{\bar\sigma 2}, \quad
   a_{{\bf k}\sigma}^{\dag} =  X\sb{i}\sp{\sigma 0} + 2\sigma
X\sb{i}\sp{2\bar\sigma } \,$  and used all   four components of the matrix GF
(\ref{s1}) $\,{\hat G}_{\alpha\beta}({\bf k},\omega)\,$    with  the diagonal
components given by (\ref{s3}). In (\ref{r10}) we introduced also the one-band
spectral functions  determined by the GFs (\ref{s9}): $\,A_{1 (2)}({\bf k},
\omega) =- (1/\pi)\,{\rm Im} {G}_{1 (2)}({\bf q},\omega) \,$. The hybridization
effects are allowed for by the parameter $\,P({\bf k})= (n-1) b({\bf k}) -   2
\sqrt{Q_1 Q_2}\,   W({\bf k})/\Lambda({\bf k})\,$.
\par
\begin{figure}[!ht]
\centering
\includegraphics[scale=.30]{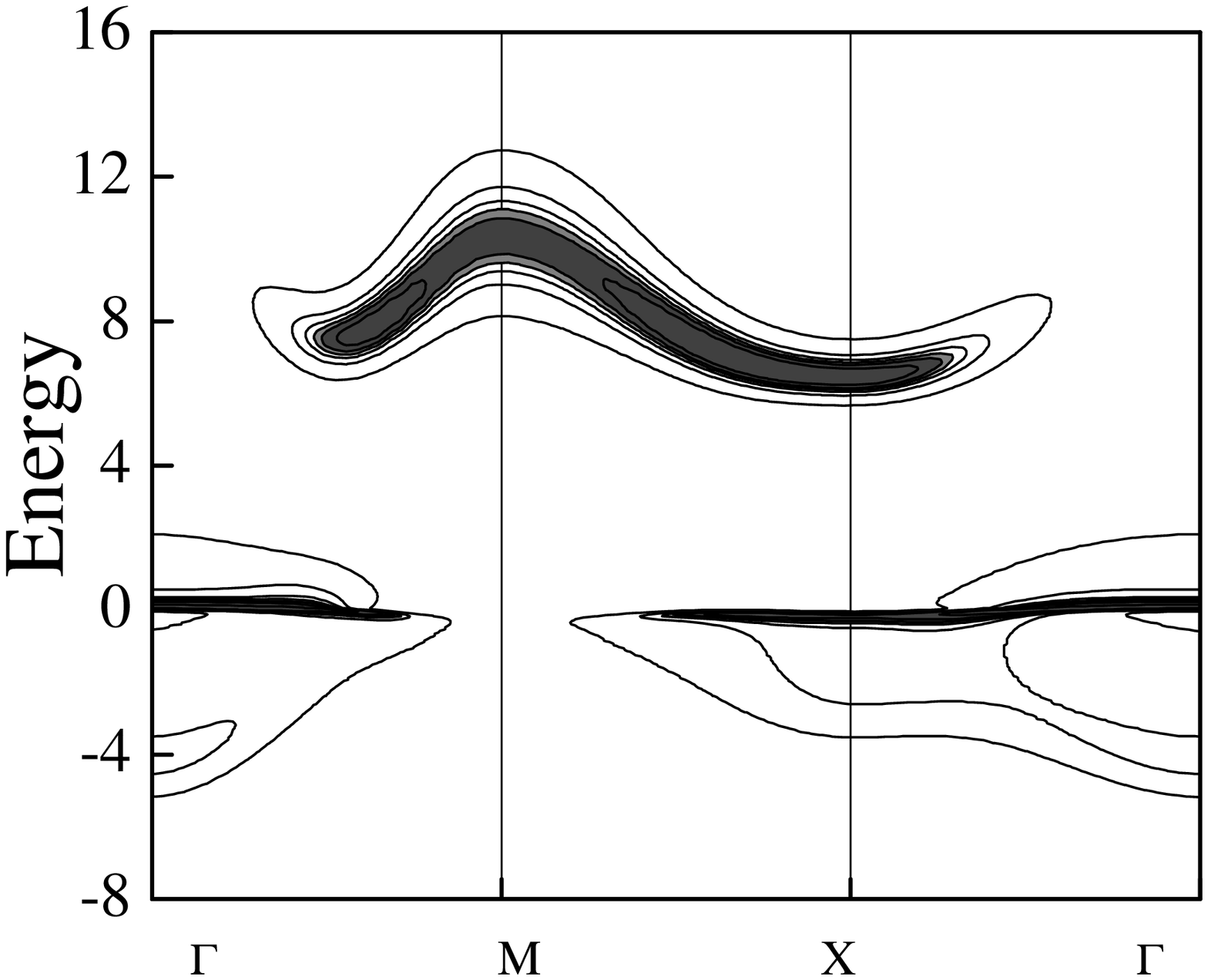}
\includegraphics[scale=.75]{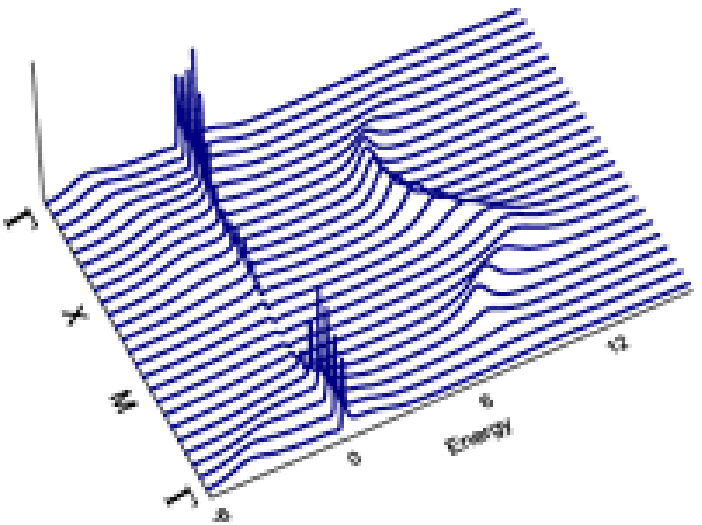}
\caption{Dispersion curves (upper panel) and  spectral functions (bottom panel)
in units of $t$ along the symmetry directions $\Gamma(0, 0)\rightarrow
M(\pi,\pi) \rightarrow X (\pi, 0) \rightarrow \Gamma(0, 0)$ for $\delta =
0.05$.} \label{figDA1-05}
\end{figure}
\begin{figure}[!ht]
\centering
\includegraphics[scale=.30]{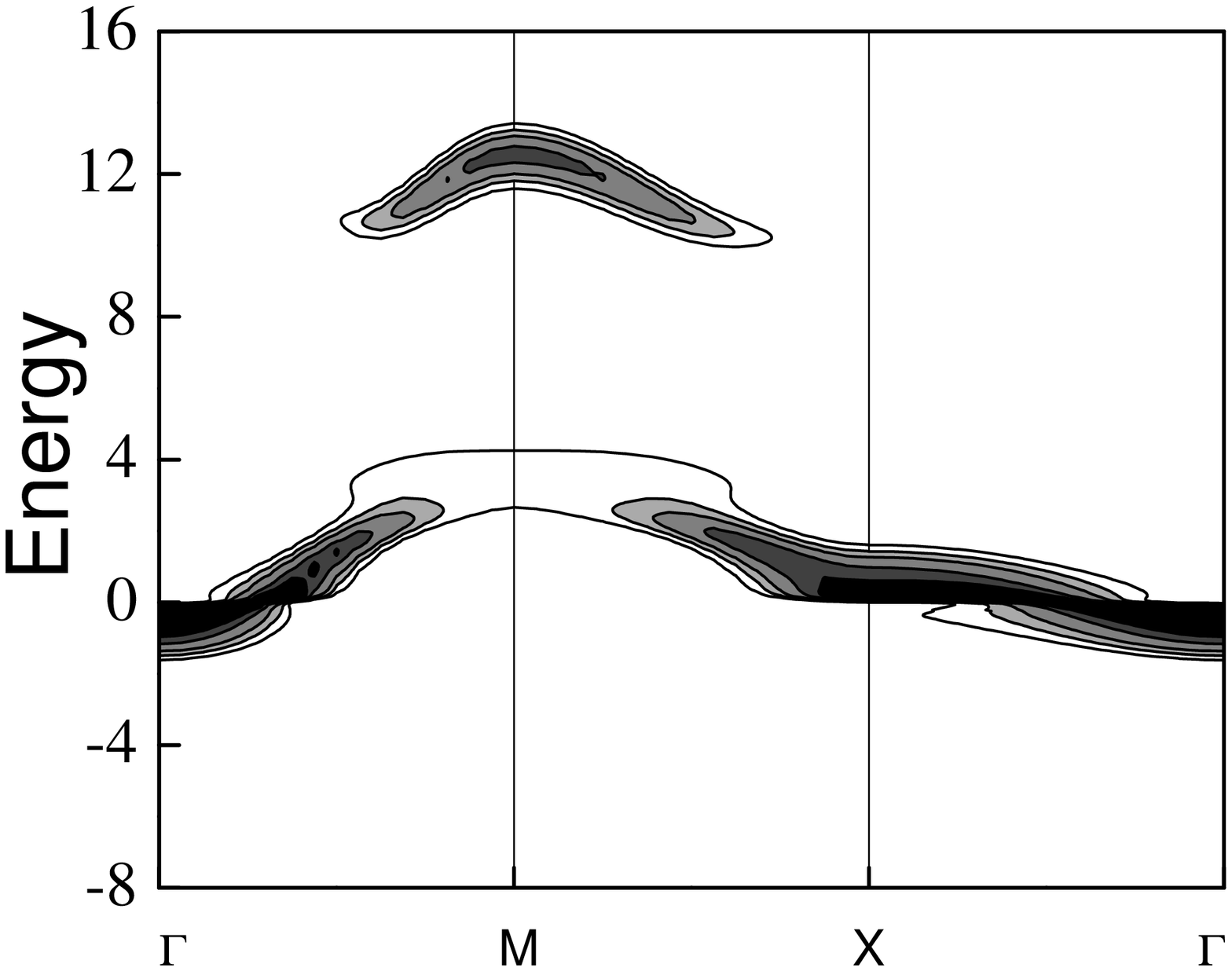}
\includegraphics[scale=.70]{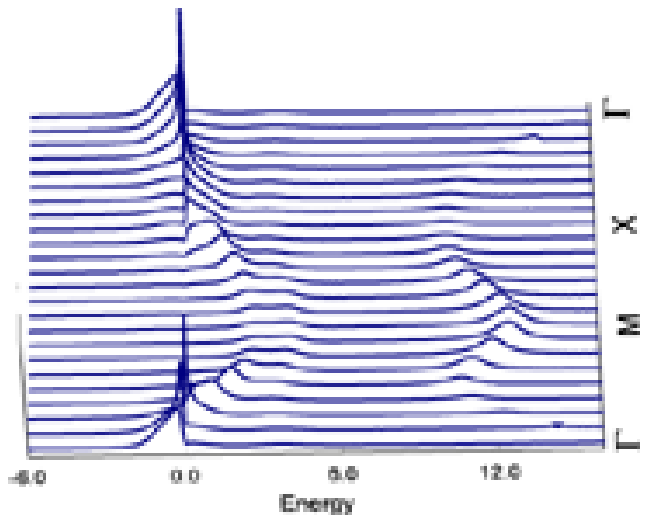}
\caption{The same as Fig.~\ref{figDA1-05} for  hole concentration $\delta =
0.3$.}
 \label{figDA1-3}
\end{figure}
\begin{figure}
\centering
\includegraphics[scale=.30]{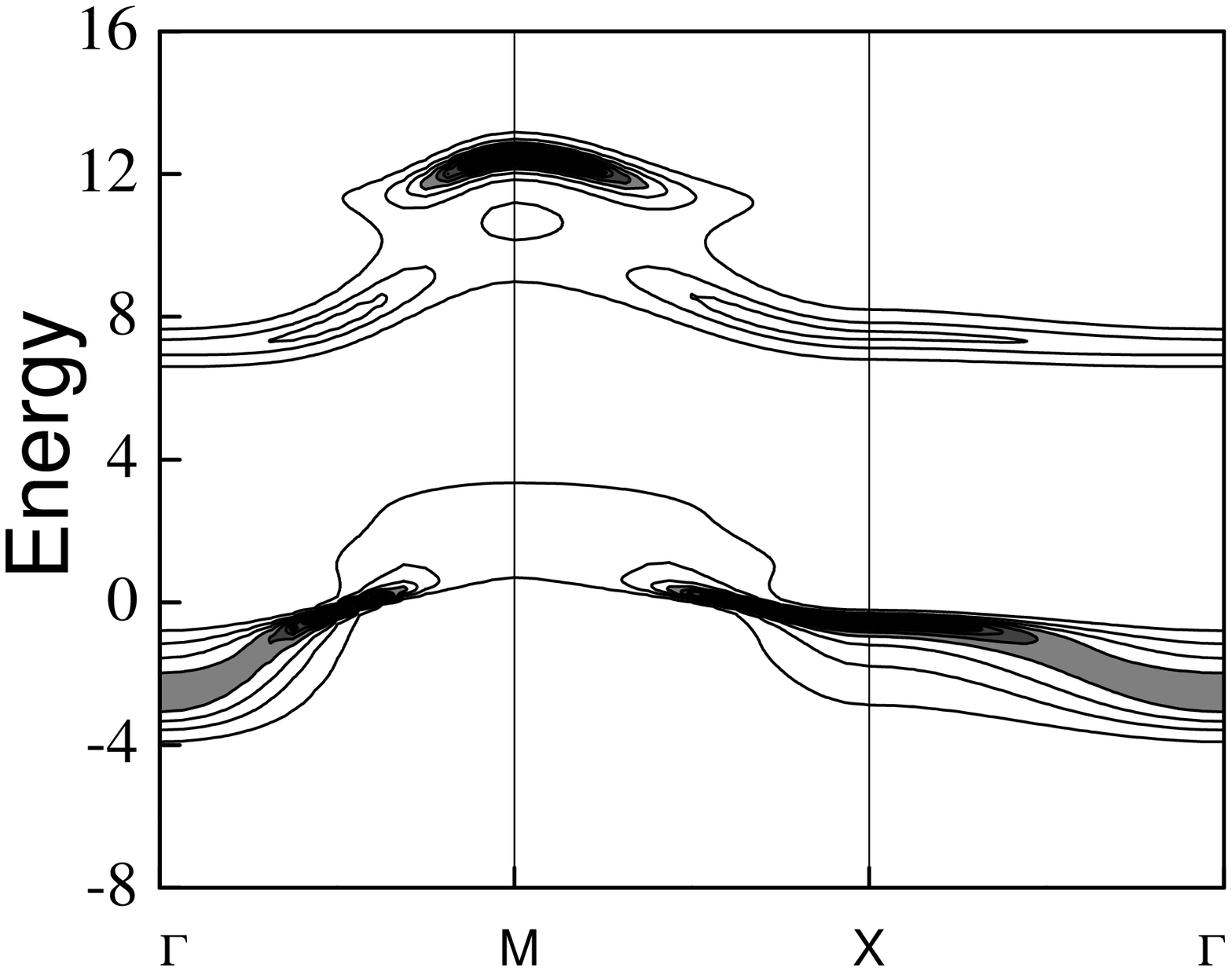}
\includegraphics[scale=.75]{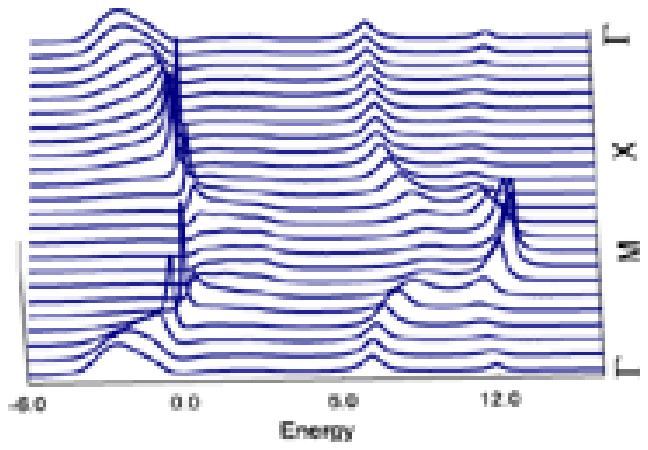}
\caption{The same as Fig.~\ref{figDA1-05} but for the hole concentration $\delta
= 0.1$ and at high temperature $T=0.3t$.}
 \label{figDA1-1T}
\end{figure}
\begin{figure}[!ht]
\centering
\includegraphics[scale=.40]{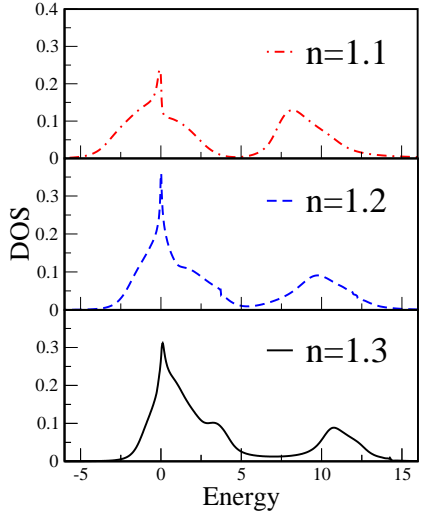}
 \caption{(Color online) Doping dependence of the electronic density of states.}
 \label{figDOS}
\end{figure}

The dispersion curves given by  maxima of  spectral functions (\ref{r10}) were
calculated for hole doping $\delta = 0.05 - 0.3$. At low hole doping, $\delta =
0.05, \, 0.1$, the dispersion reveal  a rather flat hole-doped band at the Fermi
energy (FE) ($\omega =0$) as shown in the upper panel in Fig.~\ref{figDA1-05}.
 The corresponding spectral function (the bottom panel)
demonstrates weak QP peaks at the Fermi energy.
With doping, the dispersion and  the intensity of the QP peaks at the Fermi
energy substantially increase as demonstrated in Fig.~\ref{figDA1-3} though a
flat band in $X (\pi, 0) \rightarrow \Gamma(0, 0)$ direction is still observed
in accordance with ARPES measurements in the overdoped
La$_{1.78}$Sr$_{0.22}$CuO$_4$~\cite{Yoshida01}.  To study an influence of AF
spin-correlations on the spectra, we calculate the spectral functions  at high
temperature $\,T = 0.3t\,$ for $\delta = 0.1\,$ by neglecting spin correlation
functions (\ref{n3}) in the single-particle excitation spectra (\ref{n1}) in MFA
and taking a small AF correlation length $(\xi = 1.0)$ in the
spin-susceptibility (\ref{r1}). Figure \ref{figDA1-1T} shows a strong increase
of  the dispersion and  the intensity of the QP peaks at the Fermi energy as in
the overdoped region, $\delta = 0.3 $, which proves a strong influence of AF
spin-correlations on the spectra.
 A crude estimation of the Fermi velocity from
the dispersion curve in the $\Gamma(0, 0)\rightarrow M(\pi,\pi)$ direction in
Fig.~\ref{figDA1-3} for the overdoped case  gives the value $V_{F} \simeq 7.5 t
$~\AA~$\simeq 3$~(eV$\cdot$\AA) for the hopping parameter $t = 0.4$~eV which can
be compared with experimental results $V_{F} \simeq 2.2$~(eV$\cdot$\AA) for
overdoped La$_{1.78}$Sr$_{0.22}$CuO$_4$~\cite{Yoshida01} and $V_{F} \simeq
3.9$~(eV$\cdot$\AA) for overdoped Bi-2212~\cite{Kordyuk05}.
With doping, the electronic density of states (DOS) shows a weight transfer from
the upper one-hole subband to the lower two-hole singlet subband as shown in
Fig.~\ref{figDOS}. However, even in the overdoped case a noticeable part of the
DOS retains in the upper one-hole subband.
\par
It is interesting to compare our results with those  obtained in  the
generalized DMFT~\cite{Sadovskii05} which should be close to each other as
discussed at the end of Sect.~\ref{system}.  In fact, the spectral function
shown in Fig.~8 in~\cite{Sadovskii05} for $t' = -0.4$ demonstrates a similar
flat QP bands in $\Gamma(0, 0)\rightarrow X (\pi, 0)$ and $\Gamma(0,
0)\rightarrow M(\pi,\pi) $ directions, as in our Fig.~\ref{figDA1-05} and
Fig.~\ref{figDA1-3}, a strong intensity transfer from the lower electronic
Hubbard band (LHB) to the upper Hubbard band (UHB) at the $M(\pi,\pi) $ point of
the BZ and a splitting of the LHB close to the $X (\pi, 0)$ point.  An analogous
temperature and doping ($ \xi$) behavior of  the spectral functions and the
pseudogap  revealed in the both theories supports the spin-fluctuation scenario
of the pseudogap formation. A similar behavior was observed also   in the
cluster perturbation theory~\cite{Tremblay06} (see Fig.~2~(a)
in~\cite{Senechal04}).

\subsection{Fermi surface and occupation numbers}
\label{FS}

\begin{figure}[!ht]
\centering
\includegraphics[scale=.35]{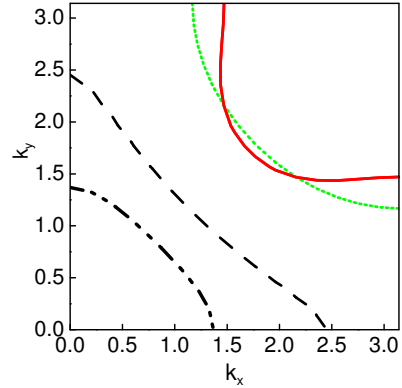}
 \caption{(Color online) Doping dependence of the  FS for  $\delta =
0.1$ (full line at $T=0.03t$ and dotted line at $T=0.3t$), $\delta = 0.2$
(dashed line), and $\delta = 0.3$ (dot-dashed line).}
 \label{figFS}
\end{figure}

\begin{figure}[!ht]
\centering
\includegraphics[scale=.20]{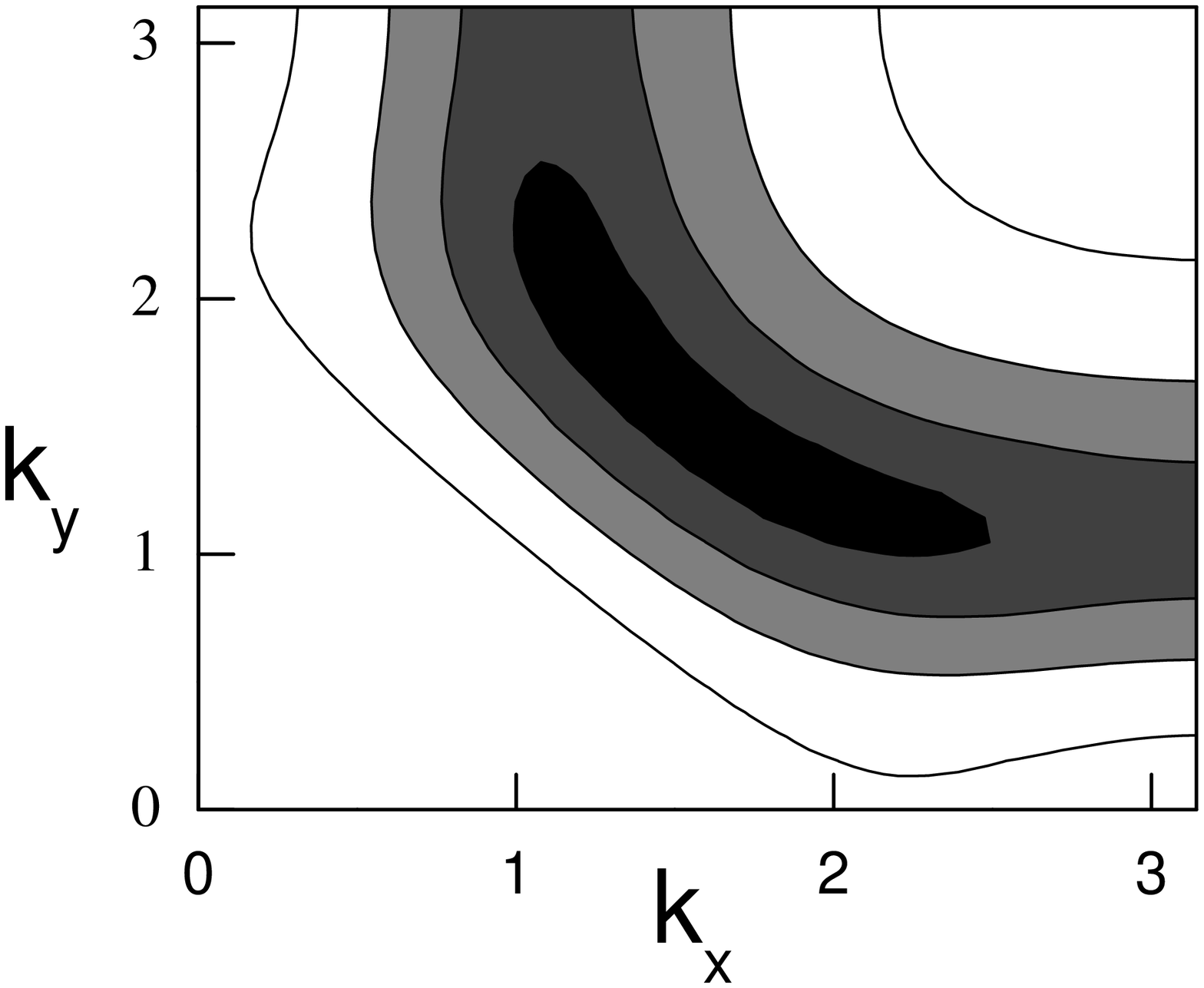}
\includegraphics[scale=.20]{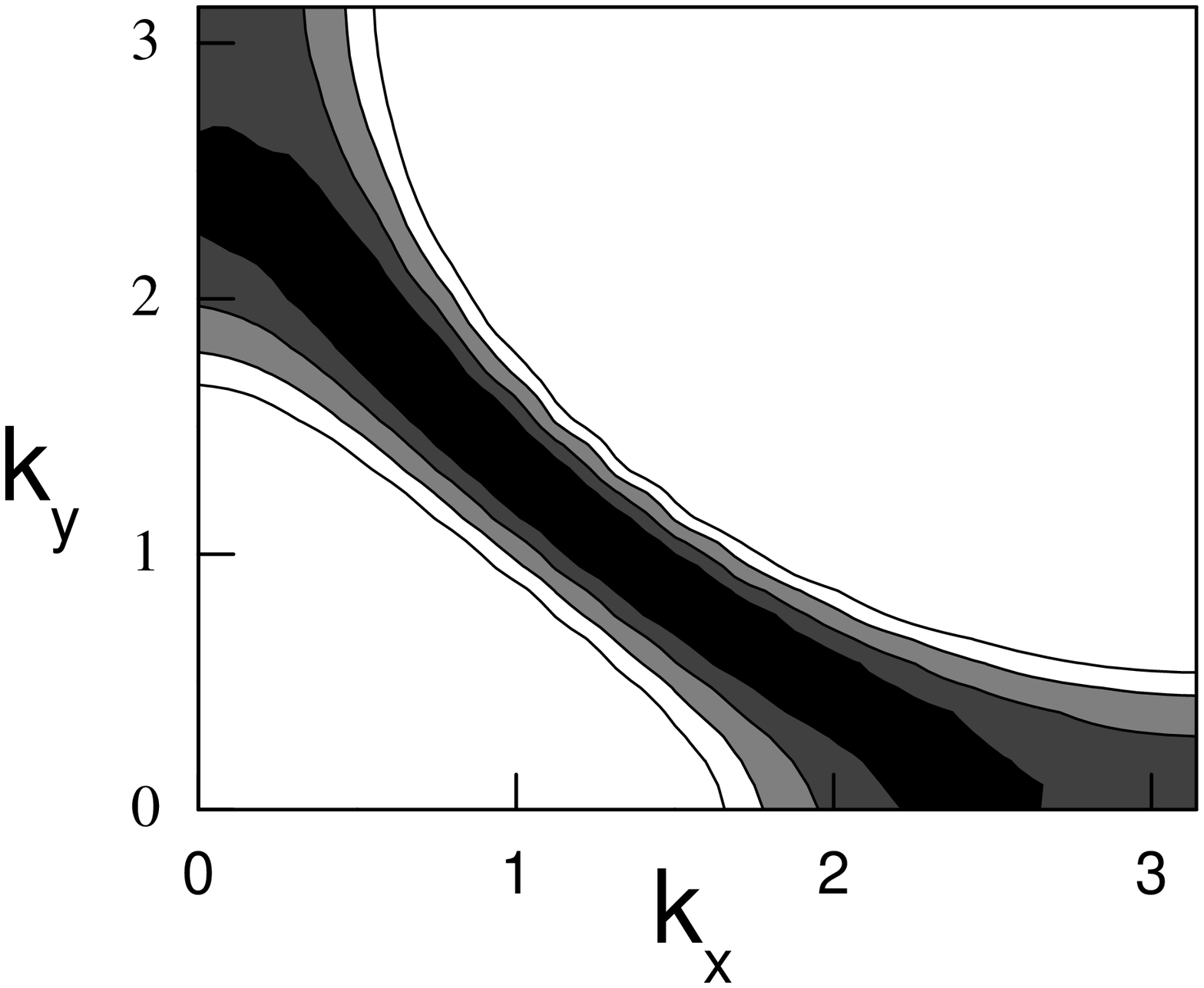}
 \caption{$A({\bf k},\omega =0)$ on the FS for $\delta = 0.1$ (left panel),
 $\delta = 0.2$ (right panel).}
 \label{figF1-2}
\end{figure}

The Fermi surface for the two-hole subband was determined by  a conventional
equation:
\begin{equation}
\varepsilon_{2}({\bf k_{\rm F}}) + {\rm Re}\,\Sigma({\bf k}_{\rm F}, \omega=0) =
0 , \label{r11}
\end{equation}
as shown in Fig.~\ref{figFS}, and then compared with those one obtained from
maxima of the spectral function $A_{el}({\bf k}, \omega = 0) $ on the $(k_x,
k_y)$-plane for $\delta = 0.1, \, 0.2$ shown in Fig.~\ref{figF1-2}. The FS
changes from a hole arc-type at $\delta = 0.1$ to an electron-like one at
$\delta =0.3$. Experimentally an electron-like FS was observed in the overdoped
La$_{1.78}$Sr$_{0.22}$CuO$_4$~\cite{Yoshida01}. The doping dependent FS
transformation can be also observed by studying the electron occupation numbers.
The electron occupation numbers in $({\bf k})$-space for one spin-direction
equal to $N_{(el)}({\sigma},{\bf k})= 1- N_{(h)}(\sigma,{\bf k})$ where the hole
occupation numbers $N_{(h)}(\sigma,{\bf k}) \equiv N_{(h)}({\bf k})$ according
to (\ref{m2}) are determined only  by the diagonal GFs (\ref{s3}). From the
latter equation and (\ref{s9}) we get:
\begin{eqnarray}
N_{(h)}({\bf k}) &=&
 [Q_1 + (n-1)b({\bf k})]\, {N}_{1}({\bf k})
\nonumber\\
 & + & [ Q_2 - (n-1)b({\bf k})\, {N}_{2}({\bf k}),
\nonumber\\
{N}_{1(2)}({\bf k})&=& -\frac{1}{\pi}\,\int^{\infty}_{-\infty}
\frac{d\omega}{e^{\omega/T} +1}\, \mbox{Im}\, { G}_{1(2)}({\bf k},\omega)
\nonumber\\
& = & \frac{1}{2}+ \frac{T}{2} \sum_{m=-\infty}^{\infty } \,
   G_{1(2)}({\bf k}, i\omega_{m}).
 \label{r12}
\end{eqnarray}
\begin{figure}[!ht]
\centering
\includegraphics[scale=.62]{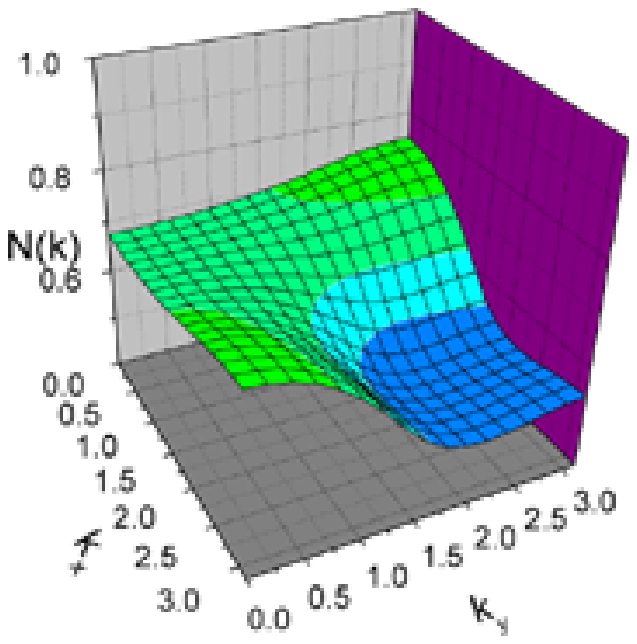}
\includegraphics[scale=.62]{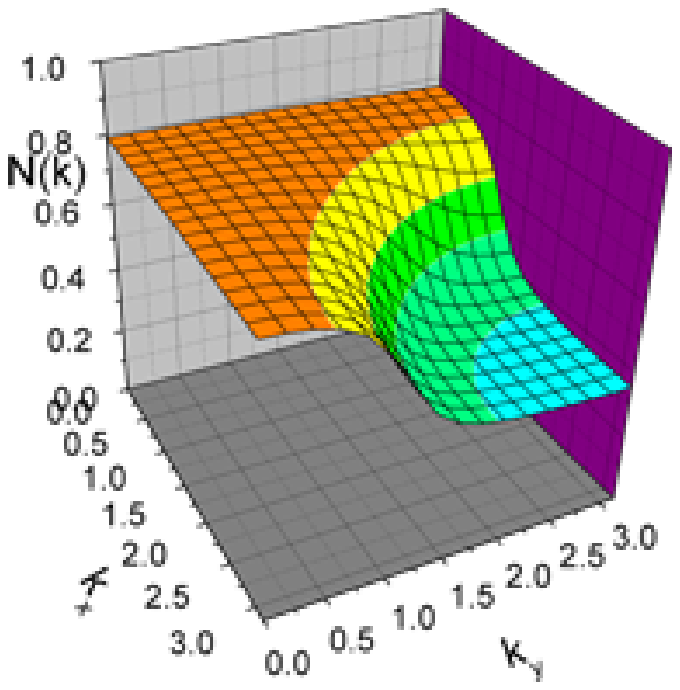}\\[0.5cm]
 \caption{(Color online) The electronic occupation numbers $N_{\bf k}$
for  $\delta = 0.1$  at $T=0.03t$ (upper panel) and  at $T=0.3t$ (bottom
panel).}
 \label{figNk1-1}
\end{figure}

\begin{figure}[!ht]
\centering
\includegraphics[scale=0.70]{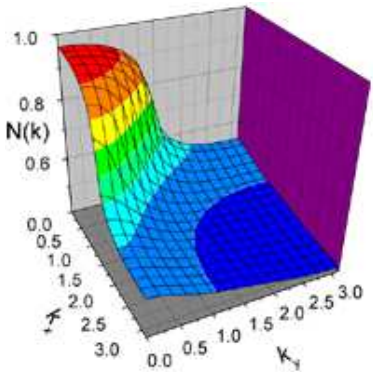}\\[0.5cm]
 \caption{(Color online) The electronic occupation numbers $N_{\bf k}$ at $T=0.03t$ for   $\delta = 0.3$.}
 \label{figNk3}
\end{figure}
The electron occupation numbers in a quarter of the BZ
 $\,(0 < k_x, k_y < \pi )\, $ are shown in Fig.~\ref{figNk1-1} for $\delta = 0.1$
at low temperature $T = 0.03t$ and   at high temperature $T = 0.3t$. With doping
the the shape of the $N_{\bf k}$  is changing revealing a transition of the
hole-like FS to the electron-like in the overdoped case $\delta = 0.3$ as
plotted in Fig.~\ref{figNk3}.
\par
While in the underdoped case at $\delta = 0.1$ the drop of the occupation
numbers at the Fermi level crossing  is rather small, $\Delta N_{(el)} \simeq
0.15$, for high temperature $T = 0.3t$ or in the overdoped case at $\delta =
0.3$ when the AF spin correlations are suppressed, the occupation number drops
are substantially increased:  $\Delta N_{(el)} \simeq 0.45, \, 0.55$,
respectively. Thus, the arc formation and a small change of the electron
occupation numbers at the FS crossing at low doping further prove  a large
contribution of the spin correlations in the renormalization of QP spectra.
\par
The obtained result concerning  the ``destruction" of the FS caused by the arc
formation shown in Fig.~\ref{figF1-2} and Fig.~\ref{figFA4} for low doping,
which corresponds to large  $ \xi $,  correlates well with the studies  within
the generalized DMFT~\cite{Kuchinskii05}. As shown in Fig.~2
in~\cite{Kuchinskii05}, the spectral density intensity plots clearly demonstrate
the arc formation on the FS  for large coupling constant $\lambda_{sf} = \Delta
= 2t$ and $ \xi = 10$, while the FS determined from (\ref{r11}) gives several
solutions as in our Fig.~\ref{figFS4} for $U_{eff} = 4\, t \,$ in
Sect.~\ref{D4}.

\subsection{Self-energy and kinks}
\label{SE}

\begin{figure}[!ht]
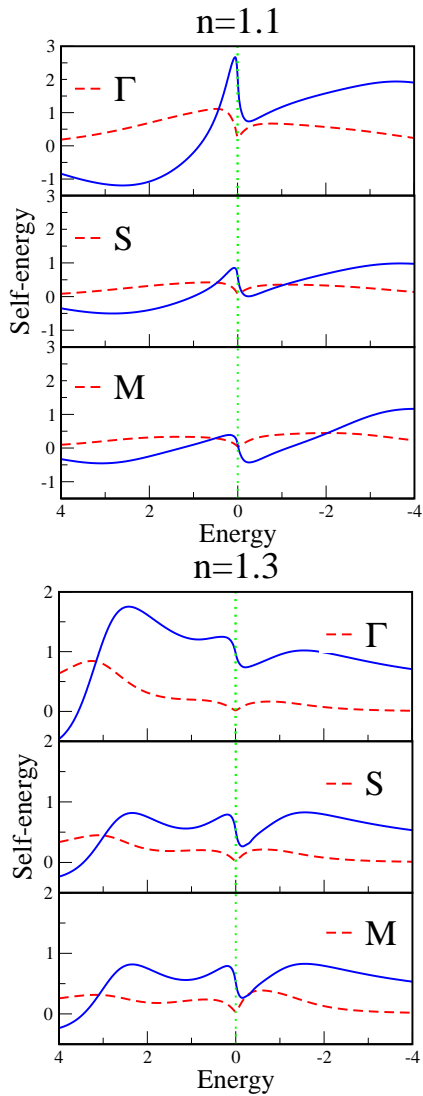

\centering
\includegraphics[scale=.40]{fig9a.eps}
\includegraphics[scale=.40]{fig9b.eps}
 \caption{(Color online) Energy dependence of the real and imaginary parts of the self-energy
$\Sigma({\bf k}, \omega)$  at the $\,\Gamma(0,0)$, $\, S(\pi/2,\pi/2)\,$ and $\,
M(\pi,\pi)\,$ points at $\delta = 0.1$ (upper panel) and $\delta = 0.3$ (bottom
panel).}
 \label{figSE-1-3}
\end{figure}

\begin{figure}[!ht]
\centering
\includegraphics[scale=.27]{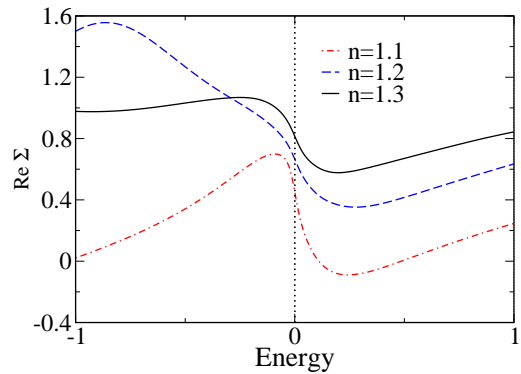}
 \caption{(Color online) ${\rm Re}\Sigma({\bf k}, \omega)$ in the
 $\Gamma(0, 0)\rightarrow M(\pi,\pi)$ direction at the FS.}
 \label{figSEn}
\end{figure}

\begin{figure}[!ht]
\centering
\includegraphics[scale=.20]{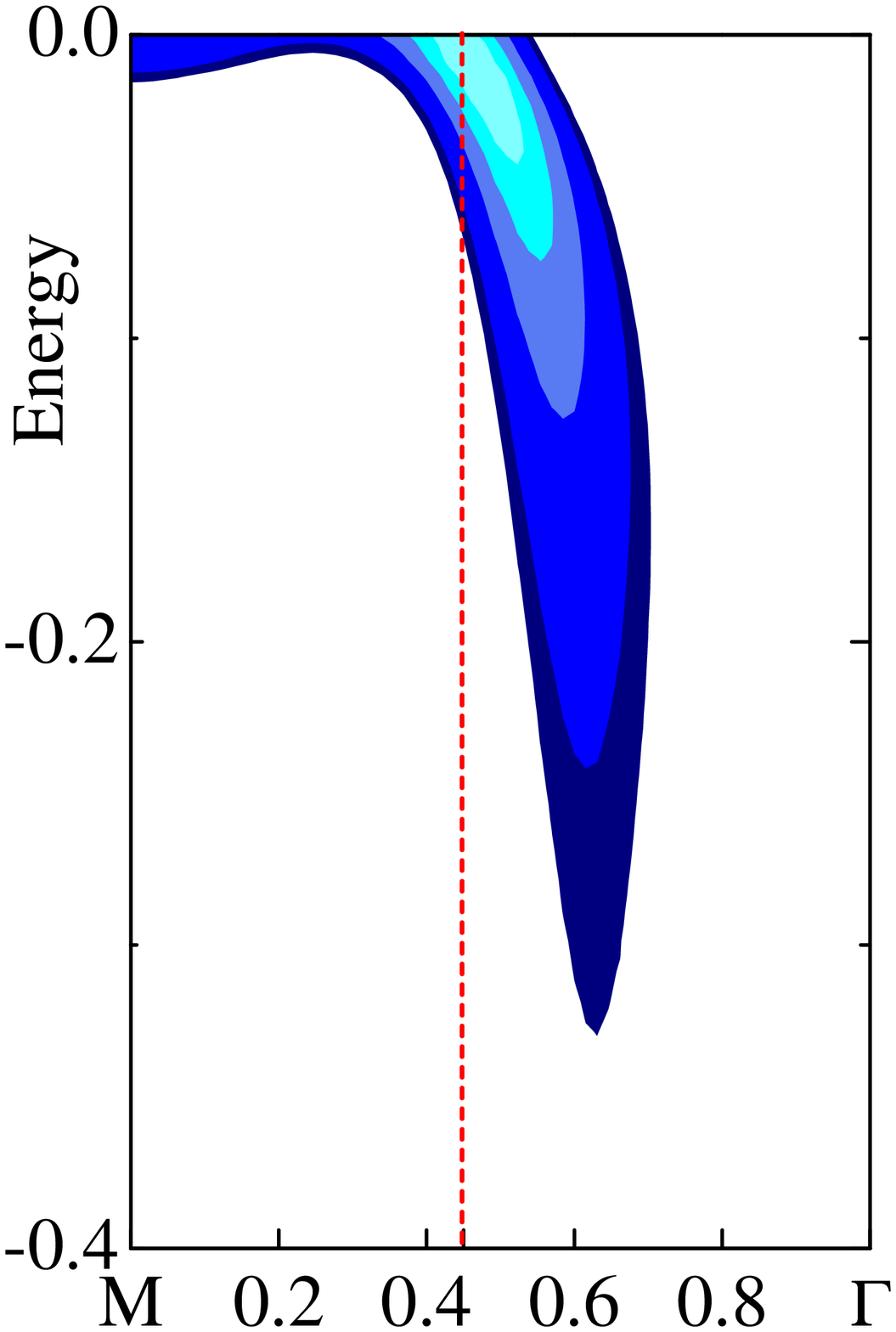}
\includegraphics[scale=.20]{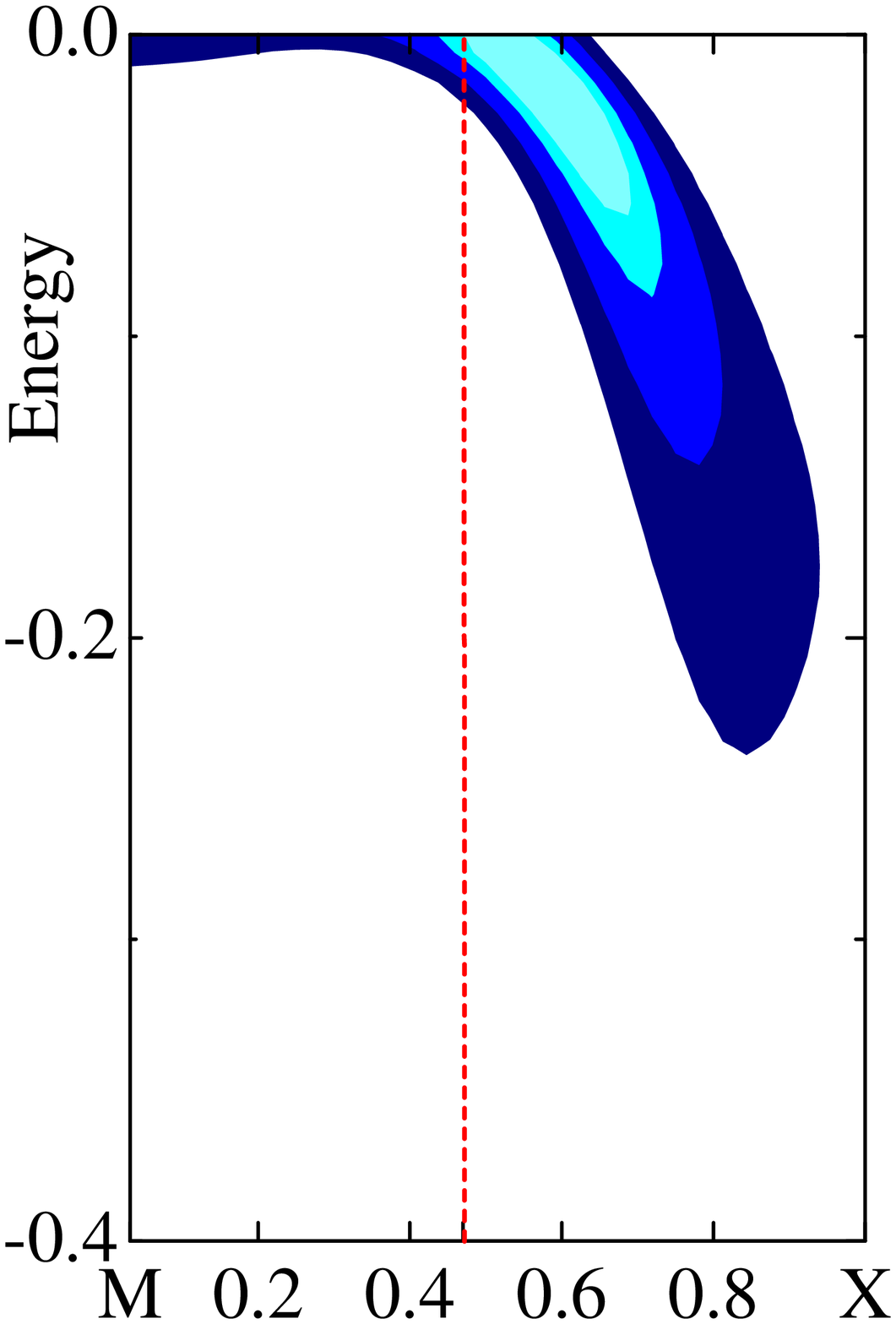}
 \caption{(Color online) Dispersion curves along the symmetry directions
 $M(\pi,\pi) \rightarrow \Gamma(0, 0) $ (left panel) and
 $M(\pi,\pi) \rightarrow X (\pi, 0)  $  (right panel) in
units of $t$ for $\delta = 0.1,\,T = 0.03 t$. Fermi level crossing is shown by
vertical dotted line.}
 \label{figK1-1}
\end{figure}

\begin{figure}[!ht]
\centering
\includegraphics[scale=.20]{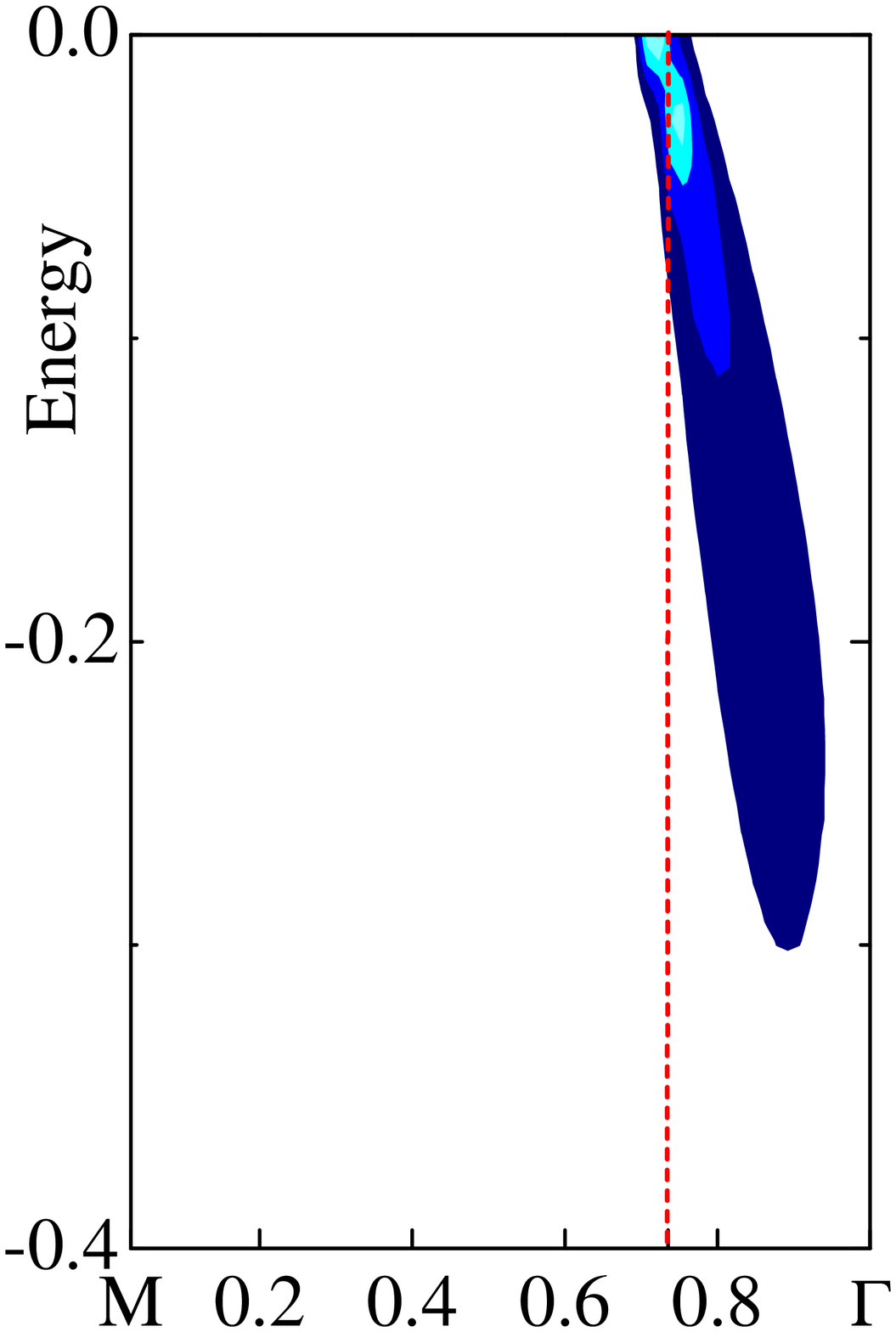}
\includegraphics[scale=.20]{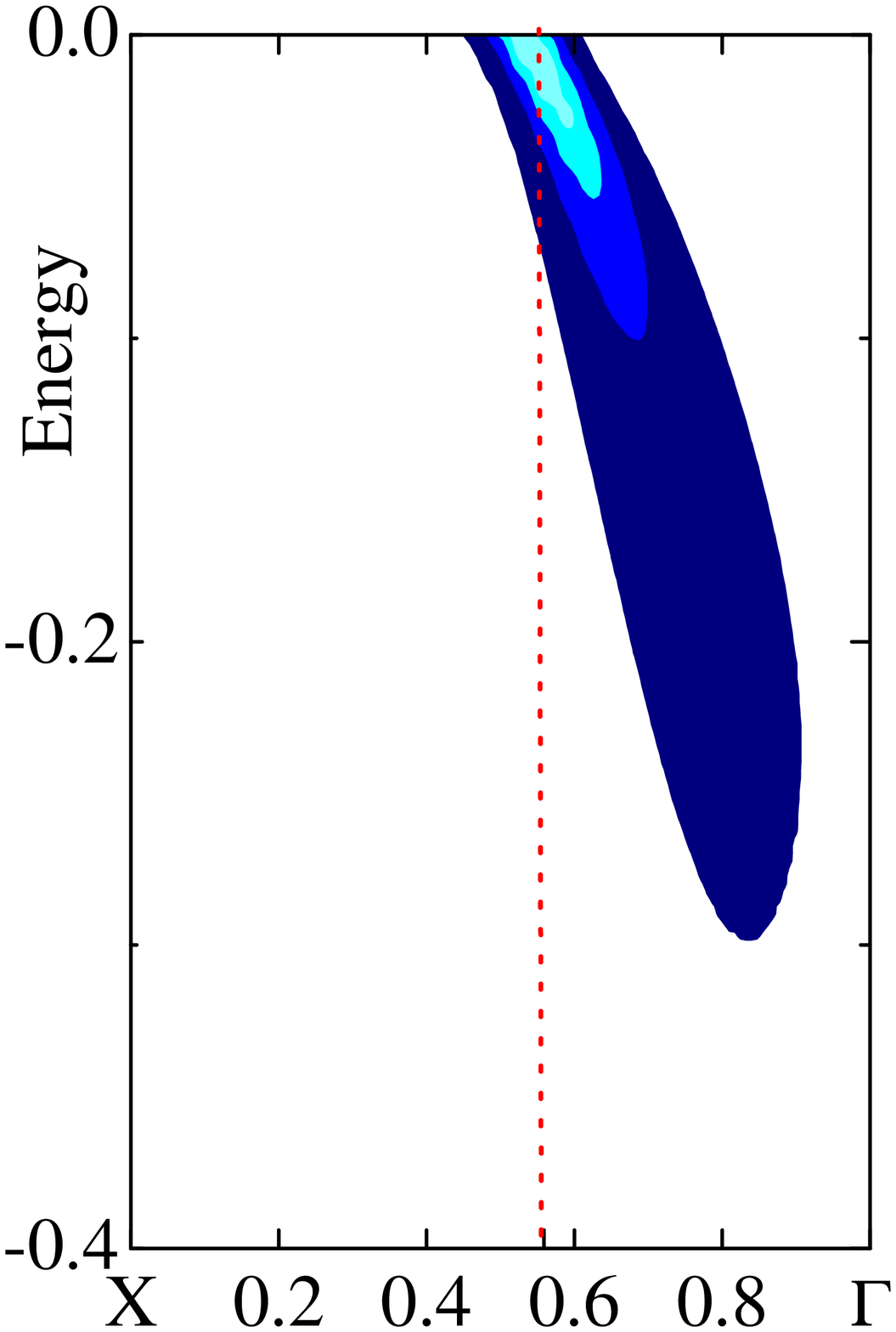}
 \caption{(Color online) The same as Fig.~\ref{figK1-1} but for $\delta = 0.3$
 along the symmetry directions  $M(\pi,\pi) \rightarrow \Gamma(0, 0) $(left panel) and
 $X (\pi, 0) \rightarrow \Gamma(0,0)$  (right panel).}
 \label{figK1-3}
\end{figure}

 Energy dependence of the real and imaginary parts of the self-energy
$\Sigma({\bf k}, \omega)$ for $\delta = 0.1,\, 0.3$  at the $\Gamma(0,0)$, $
S(\pi/2,\pi/2)$ and $ M(\pi,\pi)$ points are shown in Fig.~\ref{figSE-1-3}.
These plots demonstrate a strong dependence of the self-energy on the
wave-vector and the hole concentrations. With doping, the coupling constant
substantially decreases as seen by the decreasing of  the imaginary part and the
slope of the real part at the FS crossing which determines the coupling constant
$\lambda = - (\partial\,{\rm Re}\tilde{\Sigma}({\bf k}, \omega)/\partial
\omega)_{\omega =0}\,$. As shown in Fig.~\ref{figSEn},  the coupling constant in
the $\Gamma(0, 0)\rightarrow M(\pi,\pi)$ direction  decreases from  $\lambda
\simeq 7.86$ at $\delta = 0.1\,$ to $\lambda \simeq 3.3$ at $\delta = 0.3$.
At large binding energies (greater than the boson energy responsible for the
interaction) the self-energy effects vanish and the electron dispersion should
return to the bare value, giving a sharp bend, the so-called ``kink" in the
electron dispersion. The amplitude of the kink and the energy scale where it
occurs are related to the strength of the electron-boson interaction and the
boson energy, respectively. In ARPES the kink is observed as a changing of the
slope for an intensity plot for the spectral function $ {A}({\bf k}, \omega)$ in
a particular ${\bf k}$-wave vector direction below the Fermi level $\omega \leq
0$ (for electrons). Usually two directions are studied: the nodal $(\Gamma
\rightarrow M)$ and the antinodal $(X \rightarrow M)$ ones. Intensity plots for
the spectral function $ {A}({\bf k}, \omega)$   at $\delta = 0.1$ are shown in
Fig.~\ref{figK1-1} in the nodal direction (left panel) and the antinodal one
(right panel).   The same plots  at $\delta = 0.3$ are shown in
Fig.~\ref{figK1-3} in the nodal direction (left panel)  and  $X (\pi, 0)
\rightarrow \Gamma(0,0)$ direction (right panel).
A change of dispersion is clearly seen with increasing binding energy below the
FS shown by dotted line. For the underdoped case  the kink is larger than for
the overdoped one. A crude estimation of the strength of the kink  from the
ratio of the dispersion slope $V_{\rm F }$ close to the FS $(\omega = 0)$ to
those one  $V_{\rm F }^{0}$ at large binding energy $(\omega \sim 0.2 t)$,
$V_{\rm F }^{0}/V_{\rm F } = (1 +\lambda )$, gives the following values: $(1
+\lambda ) \sim 7.6,\, 3.5$ at   $\delta = 0.1$ for the nodal and antinodal
directions, respectively. In the overdoped case the nodal value is much smaller,
while in the the antinodal $X (\pi, 0) \rightarrow \Gamma(0,0)$ direction is
still quite large: $(1 +\lambda ) \sim 2.5$. These estimations are in accord
with the evaluation of the coupling constant $ \lambda $ from the slope of the
real part of the self-energy discussed above.
\par
It is important to stress that in our theory the self-energy effects and the
corresponding kinks are induced by the spin-fluctuation spectrum in the form of
the continuum (\ref{r1}) which at low temperature $T \sim 0.03t \ll \omega_s =
0.4t$ has a large intensity already at small energy $\omega \sim 0.03t $ and
decreases slowly up to a high energy $\omega \sim t$. In the spin-fermion model
the kink phenomenon is usually explained  by electron interaction with the
spin-resonance mode $\Omega_{\rm res} \simeq 40$~meV observed in the
superconducting state. This results in a break of the electron dispersion
("kink") at a certain energy $\omega \sim \Omega_{\rm res} + \Delta_{0}$ where
$\Delta_{0}$ is the superconducting gap (see, e.g.~\cite{Eschrig05}). In the
normal state considered in our theory  the spin-resonance mode is inessential.
Its contribution amounting only few percents of the total spin fluctuation
spectrum (\ref{r2}) should not change our results which reveal a rather strong
interaction with a smooth energy variation without any specific kink energy.

\subsection{Dispersion and FS at $\bf U_{eff} = \Delta = 4t$}
\label{D4}

The effective  Coulomb energy in the Hubbard model~(\ref{m1}) $ U_{eff} = 8t$
results in a large  charge-transfer gap  $\Delta \simeq  3$~eV for $t = 0.4$~eV
even in the overdoped case, Fig.~\ref{figDA1-3}, while  experiments point to a
smaller value   of the order of $1.5 - 2$~eV. To correct this inconsistency,  we
present in this section the results obtained for a smaller value of $ U_{eff} =
\Delta = 4t$. We also take into account the hoping parameter for the n.n.n. $\,
\pm 2 a_{x}, \pm 2 a_{y}$ sites and  fix the hoping parameter in the model
dispersion ~(\ref{m1a}) as suggested for the effective Hubbard model based on
the tight-binding fitting the LDA calculations for
La$_2$CuO$_4$~\cite{Korshunov05} as follow: $\, t' = - 0.13t, \, t'' = 0.16 t$
with $t \simeq 0.7$~eV.
\begin{figure}[!ht]
\centering
\includegraphics[scale=.32]{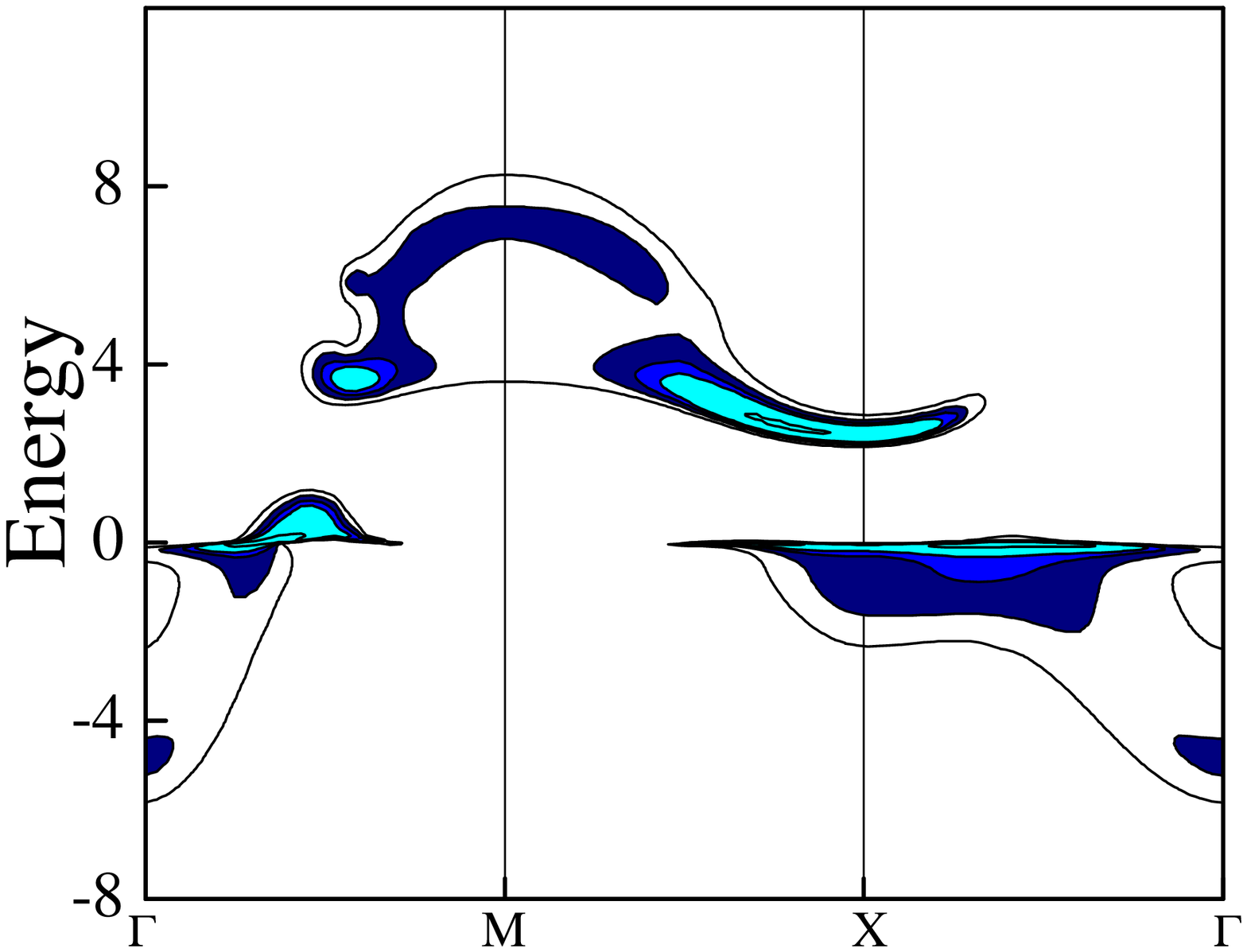}
\includegraphics[scale=.32]{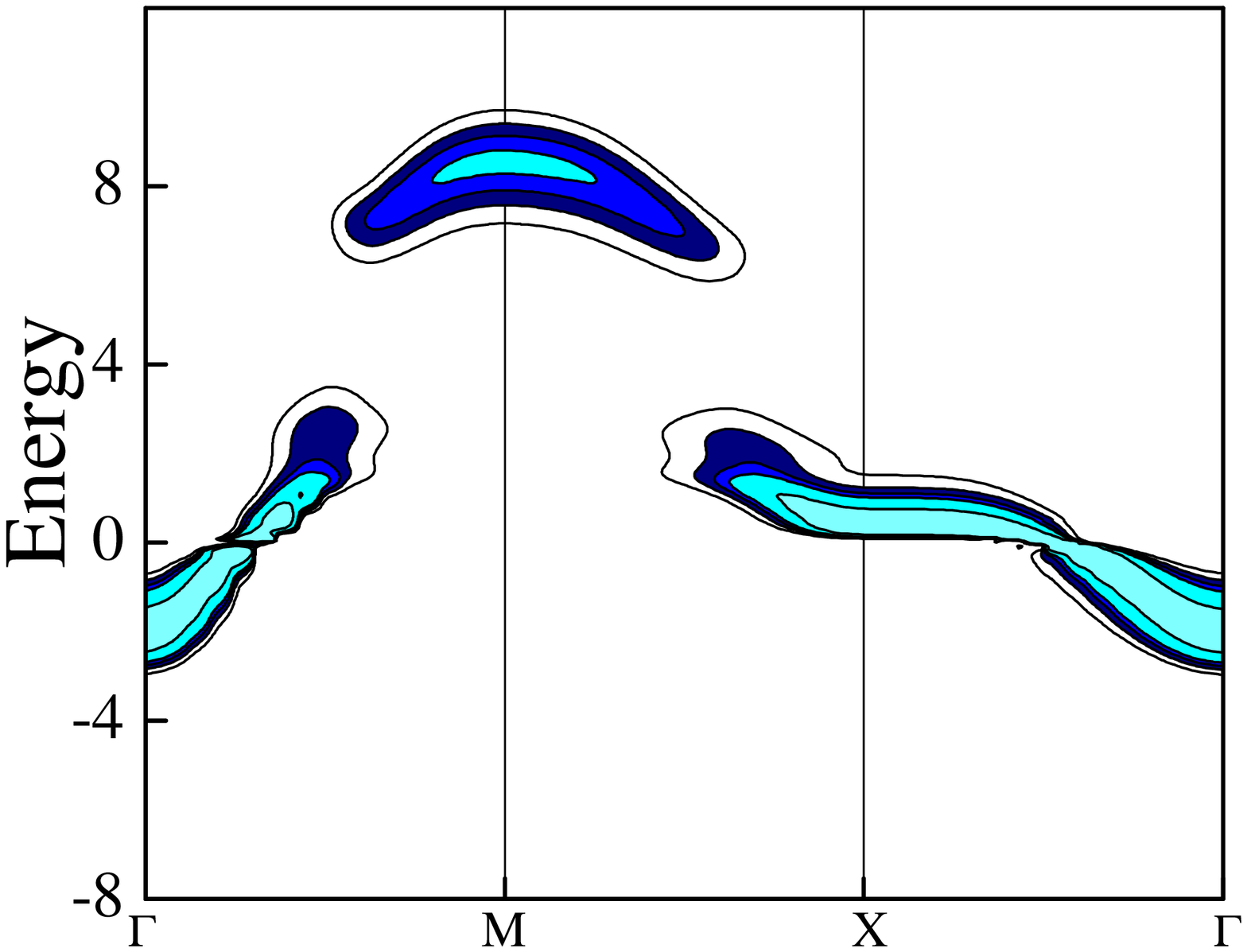}
 \caption{(Color online) Dispersion curves for $\Delta = 4t$ along the symmetry directions
$\Gamma(0, 0)\rightarrow M(\pi,\pi) \rightarrow X (\pi, 0) \rightarrow \Gamma(0,
0)$ at $\delta = 0.05$ (upper panel) and $\delta = 0.3$ (bottom panel).}
 \label{figD4-05-3}
\end{figure}
\par
\begin{figure}[!ht]
\centering
\includegraphics[scale=.40]{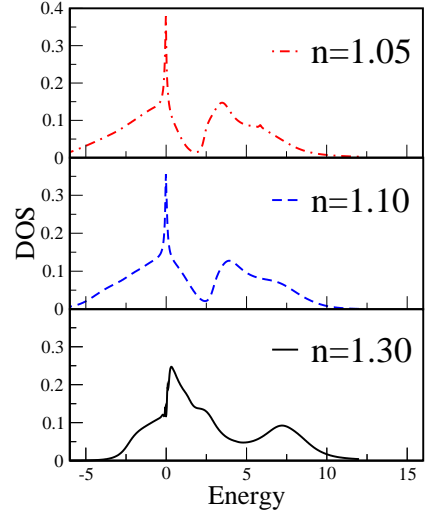}
 \caption{(Color online) Doping dependence of the DOS for $\Delta = 4t$.}
 \label{figDOS4}
\end{figure}

Main results for the dispersion and the spectral functions are not changed much
in comparison with the  previous ones as shown in Fig.~\ref{figD4-05-3}. Larger
hybridization between the subbands at small value of $ U_{eff}$ results in
increase of the dispersion and the intensity of the upper one-hole subband. This
trend is also seen in the DOS in Fig.~\ref{figDOS4}. At weak doping  the Mott
gap  between the subbands  is observed despite the intermediate Coulomb energy
$U_{eff} = 4t$, only a half of the bare bandwidth $W \simeq 8t$. This can be
explained by a reduction of the bandwidth caused by strong spin correlations in
the underdoped region up to $\tilde{W} \sim 8| t'| \,$ as discussed in
Sect.~\ref{MFA}, below the equation (\ref{n3}). In the overdoped case at $\delta
= 0.3$ when the spin correlations become weak the gap between the subbands
vanishes.
\par
\begin{figure}[!ht]
\centering
\includegraphics[scale=.35]{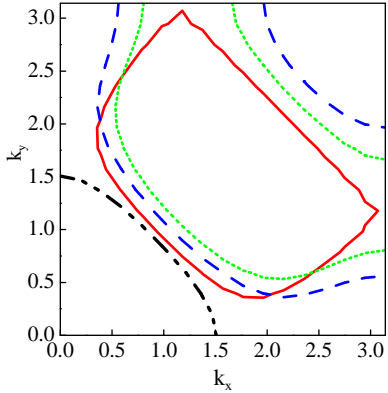}
 \caption{(Color online) Doping dependence of the  FS for  $\delta =
0.1$ (full line at $T=0.03t$ and dotted line at $T=0.3t$), $\delta = 0.2$
(dashed line), and $\delta = 0.3$ (dot-dashed line) for $\Delta = 4t$.}
 \label{figFS4}
\end{figure}
\begin{figure}[!ht]
\centering
\includegraphics[scale=.20]{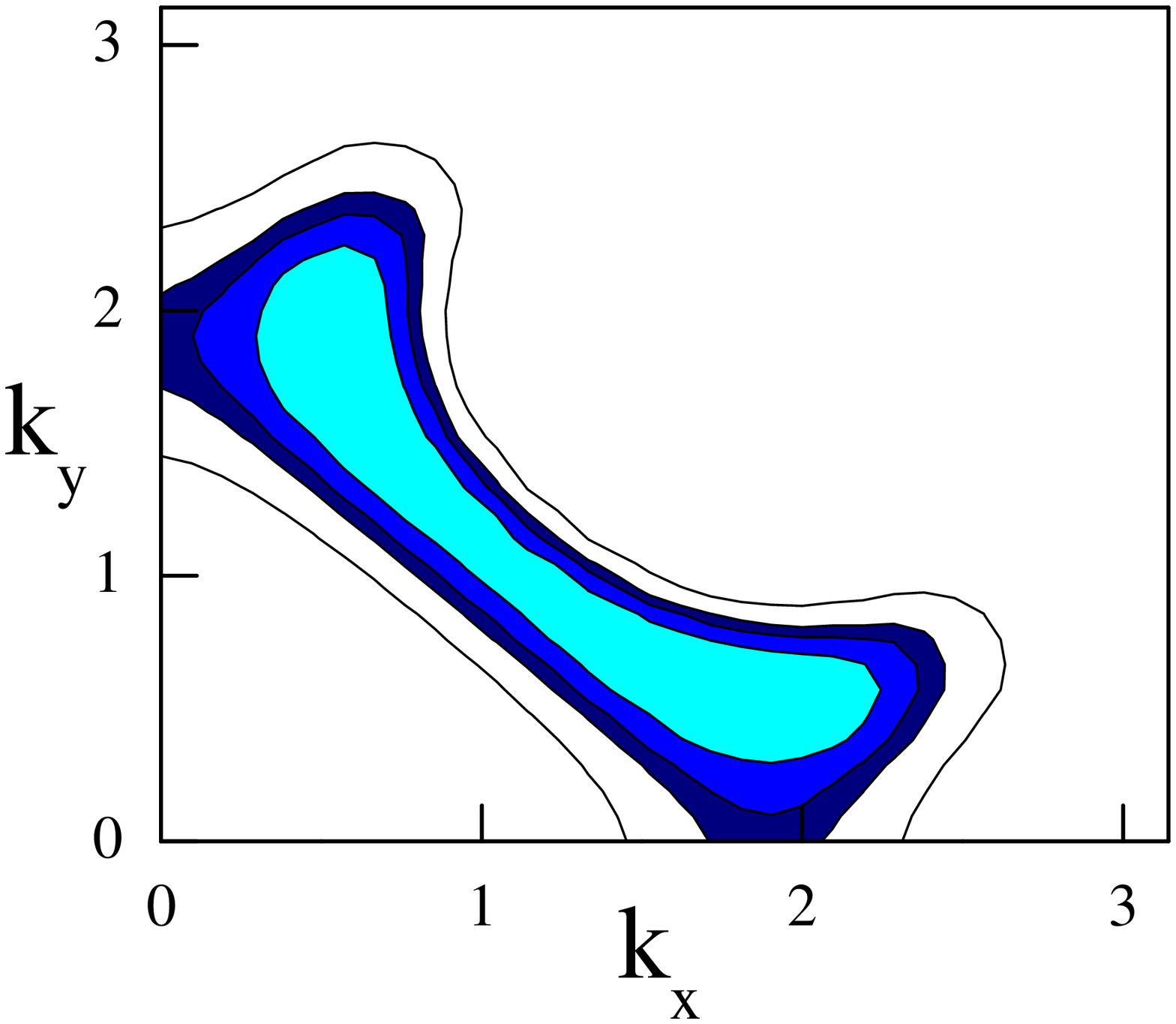}
\includegraphics[scale=.20]{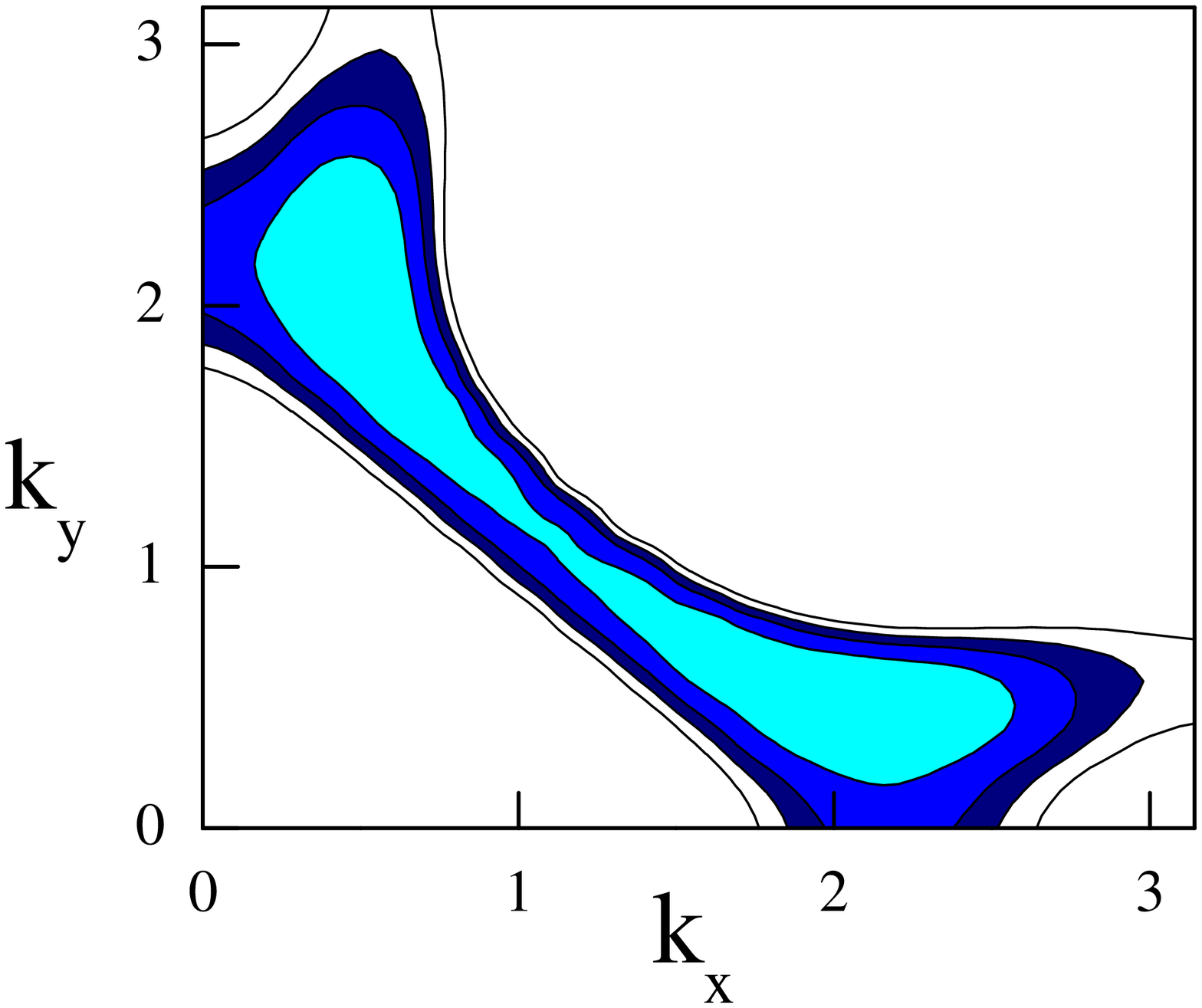}
 \caption{(Color online) $A({\bf k},\omega =0)$ on the FS at $\delta = 0.05$ (left panel) and
 $\delta = 0.1$ (right panel) at  $T=0.03t$  for $\Delta = 4t$.}
 \label{figFA4}
\end{figure}
\begin{figure}[!ht]
\centering
\includegraphics[scale=.90]{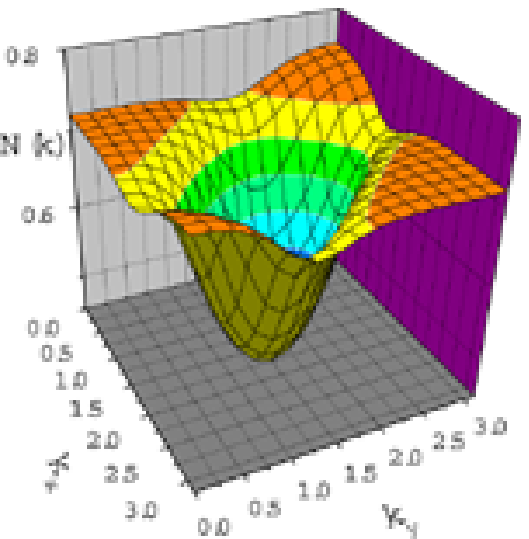}
\includegraphics[scale=.90]{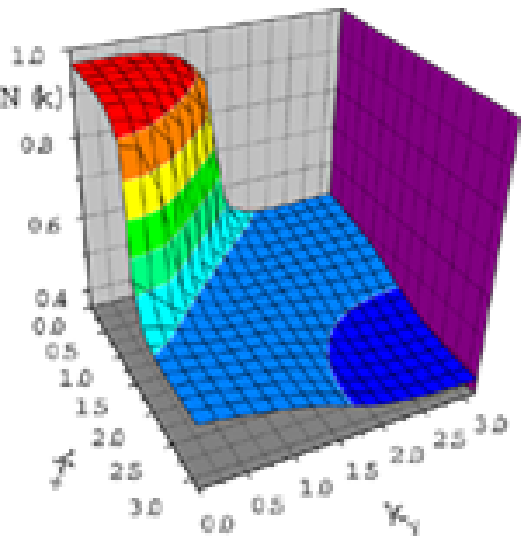}
 \caption{(Color online) The electronic occupation numbers $N_{\bf k}$ at $T=0.03t$
for  $\delta = 0.05$ (upper panel) and at $\delta = 0.3$ (bottom panel) for
$\Delta = 4t$.}
 \label{figNk4}
\end{figure}

Noticeable  changes are observed for the FS shown in Fig.~\ref{figFS4}  and in
Fig.~\ref{figFA4}. In the first plot where the FS was determined by the equation
(\ref{r11})  we see a large pocket at small doping   $\, \delta = 0.1\,$ which
opens with doping or temperature increase. At the overdoping for $\delta = 0.3$,
the FS transforms  to the electron-like as in the previous calculations. This
transformation is confirmed by calculations of the electron occupation numbers
shown in Fig.~\ref{figNk4}.
It should be noted that a pronounced hole pocket in the new set of the model
parameters is caused by the $t''$ contribution which results in a large
dispersion in the $(\pi,0) \rightarrow (0, \pi)$ direction $(\propto t''\,(\cos
2 k_x +\cos 2 k_y))$ disregarded in the previous set of the parameters. A
remarkable feature of these results  is that the part of the FS close to the
$\Gamma(0, 0)$ point in the nodal direction in Fig.~\ref{figFS4} does not shift
much with doping (or temperature) being pinned to a large FS as observed in
ARPES experiments (see, e.g.~\cite{Kordyuk05}). In fact, only this part of the
FS was detected in ARPES experiments where  the spectral function $A_{el}({\bf
k}, \omega = 0) $ shown in Fig.~\ref{figFA4} was measured.
\par
\begin{figure}[!ht]
\centering
\includegraphics[scale=.40]{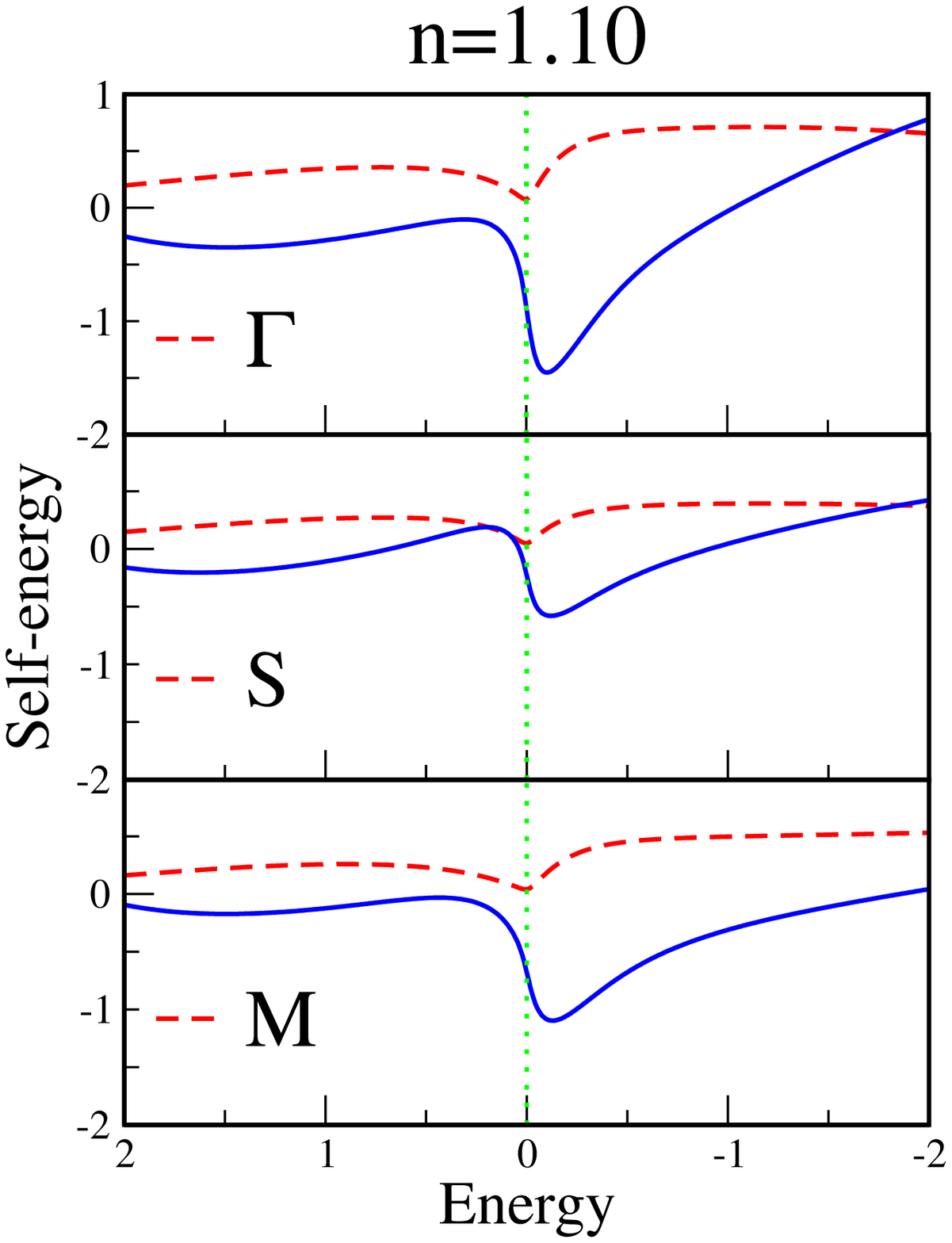}
\includegraphics[scale=.40]{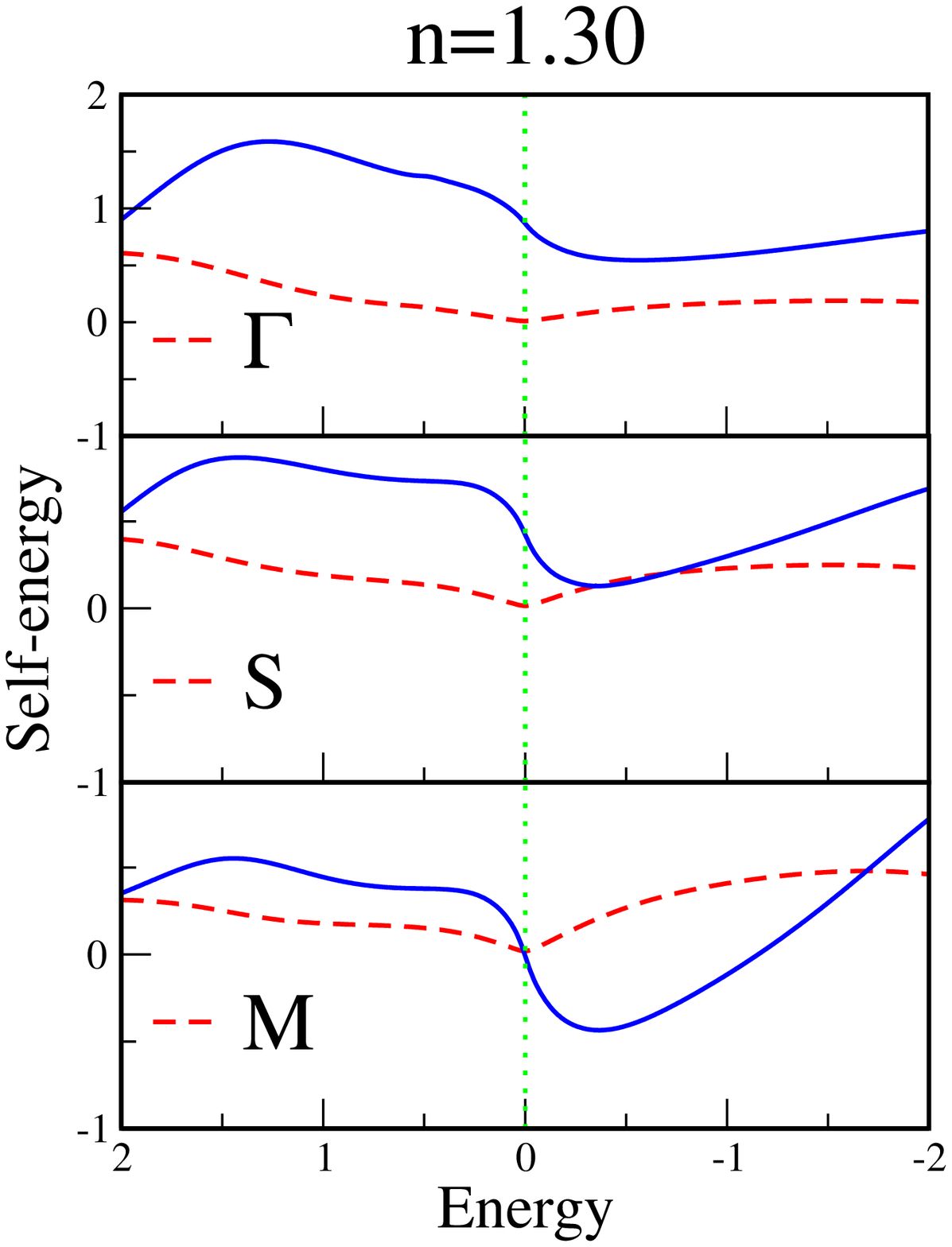}
 \caption{(Color online) Energy dependence of the real and imaginary parts of the self-energy
$\Sigma({\bf k}, \omega)$ for $\,\Delta = 4t$ at the $\Gamma(0,0)$,
$\,S(\pi/2,\pi/2)\,$ and $\, M(\pi,\pi)$ points at $\delta = 0.1$  (upper panel)
and $\delta = 0.3$  (bottom panel).}
 \label{figSE4}
\end{figure}

Concerning the self-energy effects and kinks, they are similar to the case for
$\Delta = 8t$ and confirm a strong influence of spin correlations on the QP
spectra renormalization. As shown in Fig.~\ref{figSE4}, the coupling constant
$\lambda = - (\partial\,{\rm Re}\tilde{\Sigma}({\bf k}, \omega)/\partial
\omega)_{\omega =0}\,$ being large at small doping distinctly decreases with
overdoping at $\delta = 0.3$  accompanied by suppression of the imaginary part
of the self-energy.
In conclusion, the alternative set of parameters with a  moderate effective
Coulomb energy $ U_{eff} = 4t$ in the Hubbard model ~(\ref{m1}) confirms an
important role of AF correlations in the electronic structure of system with
large single-site Coulomb interaction.

\section{Conclusion}

In the present paper the theory of electronic spectra in the strong correlation
limit for the Hubbard  model~(\ref{m1}) in a paramagnetic state has been
formulated. By employing the Mori-type projection technique for the
thermodynamic  GFs in terms of the Hubbard operators, we consistently took into
account charge carrier scattering by dynamical spin fluctuations  and derived
the self-consistent system of equations for the GF~(\ref{s9}) and the
self-energy~(\ref{s8}) evaluated in the NCA which neglects  the vertex
corrections. Though in the Hubbard model~(\ref{m1}) the electron coupling to
spin-fluctuations  is not weak, it is of the order of the hopping parameter, the
vertex corrections should not be so important in this case due to kinematic
restrictions imposed on the spin-fluctuation scattering. As was shown for the
$t$-$J$ model~\cite{Liu92}, the leading two-loop crossing diagram identically
vanishes, while the next three-loop crossing diagram gives a small contribution
to the self-energy. In any case, the NCA for the self-energy can be considered
as a starting approximation for a model with strong coupling. As we discussed at
the end of Sect.~\ref{system}, the self-consistent system of equations for the
self-energy in the classical limit  in our approach are similar to the
two-particle self-consistent approach (TPSC)~\cite{Vilk95}  or the model of
short-range static spin (charge) fluctuations~\cite{Kuchinskii06}. Numerical
results for the spectral density and the FS in the NCA approximation for the
self-energy are quite similar to the studies within the generalized
DMFT~\cite{Sadovskii05,Kuchinskii06} where all diagrams for electron scattering
by spin (or charge) fluctuations in the static approximation were taken into
account. Our results are also in accord with calculations based on the cluster
approximation~\cite{Tremblay06} and the TPSC~\cite{Vilk95}.
\par
In the present paper we have not presented  a fully self-consistent theory for
the single-electron GF and the dynamical spin and charge susceptibility. This
demands rather involved calculations of the  collective spin and charge
excitation spectra  which is beyond the scope of the present paper. Instead, we
have used a model for the dynamical spin susceptibility (\ref{r1}) which is
usually employed in phenomenological approach.  However, a variation of the
electron (hole) interaction with spin fluctuations in our theory is strongly
restricted since the vertex of the interaction is given by the hopping
parameters (\ref{m1a}) in the Hubbard model, while an intensity of spin
fluctuations at the AF wave-vector $\,{\bf  Q}\,$ ($\, C(\xi)\,$ in the
Table~\ref{Table1}) determined by the AF correlation length $\xi$  is fixed by
the sum rule (\ref{r2}). A variation of the cut-off energy $\omega_s$  does not
affect noticeably the numerical results, as we have checked. The resulting
coupling constant $\lambda$ obtained in our calculations (see Sect.~\ref{SE})
seems to be too large in comparison with ARPES results. This discrepancy can be
caused by disregarding scattering on charge fluctuations in the dynamical
susceptibility model~(\ref{s7}) and electron-phonon interaction which may reduce
the contribution from the electron-spin interaction.
\par
The main conclusion of the present study  is that a decisive role  in
renormalization of the electronic spectrum in strongly correlated system as
cuprate superconductors  is played by electron interaction with
spin-fluctuations which is in accord with other studies
(e.g.,~\cite{Eschrig05,Tremblay06,Kuchinskii06}). The numerical results for the
electron dispersion in Sect.~\ref{DA}, the FS and the occupation numbers in
Sect.~\ref{FS}, and the self-energy in Sect.~\ref{SE} unambiguously approved
this conclusion.  With doping or temperature increasing, spin correlations are
suppressed which results in transition from a strong  to a weak correlation
limit.  These observations were confirmed also by a consideration  of the model
with intermediate  Coulomb correlations in Sect.~\ref{D4}.

A theory of superconducting transition within the present theory will be
considered elsewhere.

\acknowledgments

One of the authors (N.P.) is grateful to Prof.~P.~Fulde for the hospitality
extended to him during his stay at MPIPKS, Dresden,  where a major part of the
present work has been done.

\end{document}